\documentclass[prb,showpacs,draft,amsmath,amsfonts,twocolumn,superscriptaddress]{revtex4}

\usepackage[final]{graphicx}
\usepackage{bm}
\begin{document}

\title{Time-dependent deformation functional theory}

\author{I.~V.~Tokatly}

\email{ilya.tokatly@physik.uni-erlangen.de}

\affiliation{Lerhrstuhl f\"ur Theoretische Festk\"orperphysik,
  Universit\"at Erlangen-N\"urnberg, Staudtstrasse 7/B2, 91058
  Erlangen, Germany}

\affiliation{Moscow Institute of Electronic Technology,
 Zelenograd, 124498 Russia}
\date{\today}

\begin{abstract}
We present a constructive derivation of a time-dependent deformation
functional theory -- a collective variable approach to the
nonequalibrium quantum many-body problem. It is shown that the motion
of infinitesimal fluid elements (i.~e. a set of Lagrangian trajectories) in
an interacting quantum system is governed by a closed hydrodynamics
equation with the stress force being a universal functional of the
Green's deformation tensor $g_{ij}$. Since the Lagrangian trajectories
uniquely determine the current density, this approach can be also
viewed as a representation of the time-dependent current density
functional theory. To derive the above theory we separate a
``convective'' and a ``relative'' motions of particles by
reformulating the many-body problem in a comoving Lagrangian
frame. Then we prove that a properly defined many-body wave function
(and thus any observable) in the comoving frame is a universal
functional of the deformation tensor.  Both the hydrodynamic and the
Kohn-Sham formulations of the theory are presented. In the Kohn-Sham
formulation we derive a few exact representations of the
exchange-correlation potentials, and discuss their implication for the
construction of new nonadiabatic approximations. We also discuss a
relation of the present approach to a recent continuum mechanics of
the incompressible quantum Hall liquids.

\end{abstract}


\maketitle

\section{Introduction}

Density functional theory (DFT), pioneered by Hohenberg and Kohn
\cite{HohKohn1964}, and by Kohn and Sham \cite{KohnSham1965}, is now a
standard computational tool for studying ground state properties of
various quantum many-body systems \cite{DreizlerGross1990}. By
construction, DFT, even at the exact level, yields only the ground
state energy, and the density distribution. Therefore it can not be
considered as a full alternative to the other, e.~g., field
theoretical many-body methods \cite{AGD:e}. However, the simplicity
and the computational power of DFT-based approaches makes them
practically the only tool to approach realistic quantum many particle
systems at the fully {\em ab initio} level. Moreover, from
experimental pont of view, it is frequently quite sufficient to know
just those quantities which are perfectly accessible by DFT methods.

Conceptually DFT belongs to a class of theories of collective
variables. The mapping theorems of DFT allow us to formally get rid of
the solution of the full many-body Schr\"odinger equation, and to
formulate a closed theory that operates with one, or a few basic
collective variables (the ground state density in the original
version). In this respect the equilibrium DFT is very similar to the
classical hydrostatics \cite{note0}. To find the density $n({\bf x})$
in the classical hydrostatics we need to know the equation of state,
e.~g., the dependence of pressure on the density, $P(n)$, which is
universal for a given substance. Similarly for the calculation of the
ground state density in DFT it is enough to know a universal exchange
correlation (xc) energy functional $E_{\text{xc}}[n]$. Although the
exact form of $E_{\text{xc}}[n]$ is unknown, efficient practical
approximations for it are presently available.

Away from the equilibrium the classical hydrostatics is replaced by
the classical hydrodynamics. A similar nonequilibrium extension of
DFT, a time-dependent density functional theory (TDDFT), was founded
in 1984 by Runge and Gross \cite{RunGro1984}. Although TDDFT is still
far from the maturity of its static counterpart, it has already gained
an enormous popularity in different branches of physics
\cite{TDDFT2006}. A hydrodynamic interpretation of TDDFT has been
mentioned in the original paper by Runge and Gross. However, it
remained practically out of use for more than 10 years. A comeback of
hydrodynamic analogy was caused by attempts
\cite{GroKohn1985,VigKohn1996,DobBunGro1997,VigUllCon1997} to go
beyond a simple (but formally unjustified) adiabatic local density
approximation for the time-dependent xc potential. In the mid-90th it
has been realized that nonadiabaticity of density functionals is
strongly linked to the spatial nonlocality. In particular, any
nonadiabatic xc potential must be strongly nonlocal functional of the
density. Otherwise the theory would fail to satisfy a so called
harmonic potential theorem \cite{Dobson1994,Vignale1995a}. An ultimate
connection of nonadiabaticity to the spatial nonlocality is frequently
referred to as an ultranonlocality of TDDFT.

In the linear response regime a resolution the ultranonlocality
problem was proposed by Vignale and Kohn (VK) \cite{VigKohn1996}. They
suggested to switch from TDDFT to a time-dependent current density
functional theory (TDCDFT), and to consider the current ${\bf j}({\bf
x},t)$ as a basic variable of the theory. VK have demonstrated that
the xc vector potential in linearized TDCDFT can be consistently
considered as a local functional of the current. In a subsequent paper
Vignale, Ullrich, and Conti (VUC) \cite{VigUllCon1997} have found an
elegant hydrodynamic representation of the VK result -- the time
derivative of the VK vector potential can be written as a divergence
of visco-elastic stress tensor which commonly appears in the
linearized classical continuum mechanics.  Beyond the linear response
VUC proposed an {\em ad hoc} extension of the nonadiabatic VK
functional, which simply adopts the linearized visco-elastic form of
the stress tensor for nonlinear dynamics. A similar assumption has
been made in a phenomenological approach developed by Kurzweil and
Baer \cite{KurBae2004,KurBae2005}.

A general resolution of the ultranonlocality problem in TDDFT, which
has been proposed recently \cite{TokPRB2005a,TokPRB2005b}, also
extensively relies on ideas and techniques borrowed from the classical
continuum mechanics.  Physically, the ultranonlocality is related to
the convective motion of the electron fluid (in the nonadiabatic
theory the particles at a given point of space retain the memory of
their previous positions) \cite{TokPRB2005b,TokLNP2006}.  The key idea
of Refs.~\onlinecite{TokPRB2005a,TokPRB2005b} was to eliminate the
above source of nonlocality by reformulating the theory in a
Lagrangian frame, i.e., in a local reference frame moving with the
quantum fluid. Since the convective motion in the Lagrangian frame is
absent, a spatially local description of xc effects becomes
possible. Hence the general resolution of the long standing
ultranonlocality problem can be achieved by properly changing the
"point of view": while the xc potential is extremely nonlocal in the
laboratory reference frame, it appears to be almost local from the
point of view of an observer moving with a flow. The main practical
outcome of this idea was a rigorous derivation of a nonadiabatic local
approximation for xc potential - the time-dependent local deformation
approximation (TDLDefA) \cite{TokPRB2005b,TokLNP2006}. A connection of
a general nonlinear local deformation approximation to the
phenomenological current density functional by VUC
\cite{VigUllCon1997} has been established in
Ref.~\onlinecite{UllTokPRB2006}. It turns out that the local current
density approximation by VUC corresponds to a small deformation limit
of TDLDefA. Hence the Lagrangian formulation of TDDFT can be
considered as a general framework for a description of nonadiabatic xc
memory effects. Similar to the local density approximation (LDA) in
the static DFT, the TDLDefA can serve as a basic local approximation
in the time-dependent theory.

There are, however, a few important restrictions of the formalism
developed in Refs.~\onlinecite{TokPRB2005a,TokPRB2005b}.  In these
works we reformulated the many-body theory and TDDFT in the local
comoving frame by making two simplifying assumptions: (i) we
considered a many-body system driven by a scalar external potential
(therefore an external magnetic field was excluded from the
consideration); (ii) we assumed that xc effects can also be described
by a scalar xc potential. One of the aims of the present work is to
relax both assumptions, and to formulate the theory in a most general
form.

Such a generalization is necessary, first of all, for the further
refinements and extensions of the simple local deformation
approximation.  Using the approach restricted to only scalar
potentials \cite{TokPRB2005a,TokPRB2005b} we have found that in
general the exact stress tensor in the Lagrangian frame (i.~e., in the
space seen by a comoving observer) is a functional of two collective
variables: a symmetric Green's deformation tensor $g_{ij}$, and a
skew-symmetric vorticity tensor $F_{ij}$. In the lowest order in
spatial derivatives the dependence on the vorticity disappears, and we
arrive at the local deformation approximation (TDLDefA)
\cite{TokPRB2005b,TokLNP2006}.  If we want to go beyond the local
approximation, and construct a gradient extension of TDLDefA
\cite{TaoVig2006} the dependence on $F_{ij}$ must be taken into
account, which seems to be extremely demanding technically. In the
present paper we show that this dependence is, in a certain sense,
trivial and can be singled out by reformulating the theory in terms of
vector potentials (both external and exchange-correlation). As a
result we get a theory where a properly defined stress tensor is, at
the exact level, a universal functional of only one basic variable, the
deformation tensor $g_{ij}$. As a matter of fact, this work adds one
more member to the family of nonequilibrium DFT-like theories -- a
time-dependent deformation functional theory.

Another obvious reason for reformulating the theory in terms of vector
potentials is the need to describe dynamics of many-body systems in
the presence of external magnetic fields. Recently the ideas of the
deformation-functional have been successfully applied to a
phenomenological derivation of an effective continuum mechanics of
fractional quantum Hall liquids \cite{TokPRB2006a,TokPRB2006b}.  The
present work can be considered as a formal justification (at the level
of existence theorems) of that approach.

The general idea of this paper is quite similar to that of
Ref.~\onlinecite{TokPRB2005a,TokPRB2005b}. We formulate the many-body
theory in the Lagrangian frame, and then use this formulation to
derive a closed theory of a collective variable, the time-dependent
deformation functional theory. However, the technique we develop here
is essentially different. We show that it is much more transparent and
economic to work directly with the many particle wave function, and to
formulate the problem of quantum dynamics using a Dirac-Frenkel
variational principle. Starting with the quantum mechanical action in
the laboratory frame we construct its analog in the comoving
frame. Within this approach a theory of a collective variable emerges
in a most simple and {\em constructive} fashion. We demonstrate that
the time-dependent deformation functional theory is a natural
intermediate step in solving the many-body problem in the Lagrangian
frame. Physically this can be interpreted as follows. The
transformation to the comoving frame corresponds to the separation of
the convective motion of the fluid, and the motion of particles
relatively to the convective flow. The relative motion is described by
the many-body wave function $\widetilde{\Psi}$ in the Lagrangian
space, while the convective motion is determined by a set of
trajectories ${\bf x}(\bm\xi,t)$ of infinitesimal fluid
elements. Accordingly the complete problem splits into two natural
parts. First, one solves the many-body problem in the Lagrangian
frame. This yields the wave function $\widetilde{\Psi}$ as a unique
functional of a universal geometric characteristics of the frame --
the deformation tensor $g_{ij}$ which plays a role of metric in the
Lagrangian space. On the second step we use that solution to find the
trajectories ${\bf x}(\bm\xi,t)$ from a closed hydrodynamics-like
equation. The second step is, in fact, the time-dependent deformation
functional theory, which we introduce in this paper. The theory is
formulated both in the hydrodynamic form, and in a more practical
Kohn-Sham form.

The structure of the paper is the following. In Sec.~II we illustrate
the general formalism using a pedagogical exactly solvable example of
one particle quantum dynamics. In this case the quantum problem for
the relative motion possesses an analytic solution, and the final
time-dependent deformation functional theory is formulated in an
explicit form. In the concluding subsection (Sec.~IIC) we extensively
discuss the main ideas and the results of Sec.~II. This subsection is
aimed at preparing a reader to the most general formulation of the
theory given in Sec.III. In Sec.~IIIA the Dirac-Frenkel variational
principle is used to formulate the general many-body problem in the
Lagrangian frame. The solution of the quantum problem for the relative
motion is analyzed in Sec.~IIIB. We prove the basic mapping theorem
which states that the many-body wave function in the Lagrangian frame
is a universal functional of the deformation tensor. This theorem
forms a basis of the time-dependent deformation functional theory that
can also be interpreted as an exact quantum continuum mechanics. In
Sec.~IIIB the Keldysh-contour formalism is employed to derive a closed
variational formulation of the theory. We introduce a universal
functional $W[g_{ij}^{C}]$ that plays a role of an effective
``elastic'' action of the exact quantum continuum mechanics. A
relation of this formulation to a recent magneto-elasticity theory of
fractional quantum Hall liquid \cite{TokPRB2006a,TokPRB2006b} is
discussed. In Sec~IV we present a Kohn-Sham formulation of the
theory. We introduce xc potentials both in the Lagrangian and in the
laboratory frame. We also derive a few exact representations of the xc
potentials in the laboratory frame (since just these potentials are of
practical interest). Finally, in Sec.~V we present our conclusions.

\section{Getting an idea: Quantum particle in the Lagrangian frame}

\subsection{Quantum mechanics in a local noninertial frame}

To illustrate main ideas, and the structure of a general theory
developed in Sec.~III it is instructive to consider first the simplest
case of one quantum particle moving in the presence of external
vector and scalar potentials, ${\bf A}({\bf x},t)$ and $U({\bf
  x},t)$. The system is described by the one particle wave function
$\Psi({\bf x},t)$ that satisfies the time-dependent Schr\"odingen
equation supplemented with the proper initial condition
\begin{equation} 
i\partial_{t}\Psi({\bf x},t) = H\Psi({\bf x},t), \quad
\Psi({\bf x},0) = \Psi_{0}({\bf x}),
\label{1}
\end{equation}
where $H$ is the usual one particle Hamiltonian
\begin{equation} 
H = \frac{1}{2m}[-i\partial_{\bf x} - {\bf A}({\bf x},t)]^{2} 
+ U({\bf x},t).
\label{2}
\end{equation}

For our purpose it is convenient to reformulate the problem of quantum
dynamics using a Dirac-Frenkel variational principle. The
Schr\"odinger equation, Eq.~(\ref{1}), corresponds to the condition for
the extremum of the action $S[\Psi^{*},\Psi]=\int_{0}^{t_{f}}{\cal L}dt$
with the following Lagrangian
\begin{equation}
{\cal L}[\Psi^{*},\Psi] = \int\left[
i\Psi^{*}\stackrel{\leftrightarrow}{\partial_{t}}\Psi 
- \Psi^{*}H\Psi\right] d{\bf x},
\label{3}
\end{equation}
where $\stackrel{\leftrightarrow}{\partial_{t}}$ is a symmetrized time
derivative \cite{note1}: 
$\Psi^{*}\stackrel{\leftrightarrow}{\partial_{t}}\Psi = 
(\Psi^{*}{\partial_{t}}\Psi - \Psi{\partial_{t}}\Psi^{*})/2$. 
Using Eq.~(\ref{2}) we represent the Lagrangian, Eq.~(\ref{3}), in the
following explicit form \cite{note2}
\begin{eqnarray}  \nonumber
{\cal L} &=& \int d{\bf x}\Big\{
i\Psi^{*}\stackrel{\leftrightarrow}{\partial_{t}}\Psi
- U\Psi^{*}\Psi\\
&-& \frac{1}{2m}\left[(i\partial_{x^{i}} - A_{i})\Psi^{*}\right]
\left[(-i\partial_{x^{i}} - A_{i})\Psi\right]
\Big\}.
\label{4}
\end{eqnarray}

Let us transform the equation of motion, Eq.~(\ref{1}), or,
equivalently, the Lagrangian of Eq.~(\ref{4}) to
a local noninertial reference frame moving with a given velocity 
${\bf v}({\bf x},t)$. Formally this corresponds to a nonlinear
transformation of coordinates ${\bf x}\to\bm\xi$, ${\bf x} = 
{\bf x}(\bm\xi,t)$, where the function ${\bf x}(\bm\xi,t)$ is a
solution to the following Cauchy problem 
\begin{equation} 
\frac{\partial {\bf x}(\bm\xi,t)}{\partial t} 
= {\bf v}({\bf x}(\bm\xi,t),t), \quad {\bf x}(\bm\xi,0) = \bm\xi.
\label{5}
\end{equation}
Intuitively the function ${\bf x}(\bm\xi,t)$
can be viewed as a trajectory of a small element of a fluid with the
velocity distribution ${\bf v}({\bf x},t)$. Accordingly the new
spatial coordinate $\bm\xi$ has a meaning of the initial point of that
trajectory. 

The transformation of coordinates ${\bf x}\to\bm\xi$ leads to the
following replacements in the Lagrangian
\begin{equation}
d{\bf x}\to \sqrt{g}d\bm\xi, \quad 
\partial_{x^{i}} \to 
\frac{\partial\xi^{j}}{\partial x^{i}}\partial_{\xi^{j}}, \quad
\partial_{t} \to \partial_{t} - \widetilde{v}^{i}\partial_{\xi^{j}}
\label{6}
\end{equation}
where $g(\bm\xi,t)=\det g_{ij}$ is the determinant of the induced metric 
tensor  
\begin{equation} 
g_{ij}(\bm\xi,t) = \frac{\partial x^{k}}{\partial\xi^{i}}
\frac{\partial x^{k}}{\partial\xi^{j}}; \quad
[g_{ij}]^{-1} = g^{ij} = \frac{\partial\xi^{i}}{\partial x^{k}}
\frac{\partial\xi^{j}}{\partial x^{k}}
\label{7}
\end{equation}
and $\widetilde{\bf v}(\bm\xi,t)$ is the velocity vector transformed
the moving frame:
\begin{equation} 
\widetilde{v}^{i}(\bm\xi,t) = \frac{\partial\xi^{i}}{\partial x^{j}}
v^{j}({\bf x}(\bm\xi,t),t).
\label{8}
\end{equation}
Substituting Eq.~(\ref{6}) into Eq.~(\ref{4}), and using the definitions of
Eqs.~(\ref{7}) and (\ref{8}) we obtain the following transformed
Lagrangian
\begin{eqnarray}  \nonumber
&&{\cal L} = \int \sqrt{g}d{\bm\xi}\Big\{
i\Psi^{*}\stackrel{\leftrightarrow}{\partial_{t}}\Psi
+ \Big[\frac{m}{2}\widetilde{v}^{i}\widetilde{v}_{i} + 
\widetilde{v}^{i}\widetilde{A}_{i} - U\Big]\Psi^{*}\Psi\\
&-& \frac{g^{ij}}{2m}
[(i\partial_{\xi^{i}} - \widetilde{A}_{i} -m\widetilde{v}_{i})\Psi^{*}]
[(-i\partial_{\xi^{j}} - \widetilde{A}_{j} - m\widetilde{v}_{j})\Psi]
\Big\}
\label{9}
\end{eqnarray}
Here $\widetilde{A}_{i}(\bm\xi,t)$ is the external vector potential in the
$\bm\xi$-frame:
\begin{equation} 
\widetilde{A}_{i}(\bm\xi,t) = \frac{\partial x^{j}}{\partial\xi^{i}}
A_{j}({\bf x}(\bm\xi,t),t).
\label{10}
\end{equation}
(Raising and lowering of tensor indexes are performed according to
the usual rules,
e.~g., $\widetilde{v}_{i}=g_{ij}\widetilde{v}^{j}$.) In the derivation
of Eq.~(\ref{9}) we regrouped terms to obtain a
physically expected form of $\widetilde{A}_{i}+m\widetilde{v}_{i}$ in
the kinetic energy. 
 
Apparently the second term in Eq.~(\ref{9}) plays a role of an
effective scalar potential in the moving reference frame. An important
observation is that this term exactly coincides with the classical Lagrangian
of a particle moving along the trajectory ${\bf x}(\bm\xi,t)$ in the
presence of the external fields ${\bf A}({\bf x},t)$ and $U({\bf
 x},t)$. Indeed, using Eqs.~(\ref{8}) and (\ref{10}) we find
\begin{eqnarray}\nonumber 
L^{\text{cl}}(\bm\xi,t) &=& \frac{m}{2}\widetilde{v}^{i}\widetilde{v}_{i} + 
\widetilde{v}^{i}\widetilde{A}_{i} - U\\ 
&=&\frac{m}{2}(\dot{\bf x}(t))^{2} 
+ \dot{\bf x}(t){\bf A}({\bf x}(t),t) - U({\bf x}(t),t),
\label{11}
\end{eqnarray}
where dot denotes the time derivative. Obviously the classical
Lagrangian of Eq.~(\ref{11}) parametrically depends on the initial
point, $\bm\xi$, of the trajectory.

The Lagrangian of Eq.~(\ref{10}) can be simplified by introducing
a transformed wave function $\widetilde{\Psi}(\bm\xi,t)$:
\begin{equation}
\Psi({\bf x}(\bm\xi,t),t) 
= g^{-\frac{1}{4}}e^{iS_{\text{cl}}(\bm\xi,t)}\widetilde{\Psi}(\bm\xi,t),
\label{12}
\end{equation}
where $S_{\text{cl}}(\bm\xi,t)$ is the classical action that is related the
Lagrangian $L^{\text{cl}}$, Eq.~(\ref{11}):
\begin{equation}
S^{\text{cl}}(\bm\xi,t) = \int_{0}^{t}L^{\text{cl}}(\bm\xi,t')dt'.
\label{13}
\end{equation}
The renormalization factor $g^{-\frac{1}{4}}$ in Eq.~(\ref{12})
accounts for a local change of volume induced by the nonlinear
transformation of coordinates. This allows us to preserve the
interpretation of the transformed function
$\widetilde{\Psi}(\bm\xi,t)$ as a probability density in
$\bm\xi$-space \cite{Podolsky1928,TokPRB2005a}. The exponential
prefactor in Eq.~(\ref{12}) gauges out the effective scalar potential
in  Eq.~(\ref{9}).
Inserting Eq.~(\ref{12}) into Eq.~(\ref{9}) we reduce the Lagrangian
to the following compact form
\begin{eqnarray}  \nonumber
&&{\cal L}[\widetilde{\Psi}^{*},\widetilde{\Psi}] 
= \int d{\bm\xi}\Big\{
i\widetilde{\Psi}^{*}\stackrel{\leftrightarrow}{\partial_{t}}
\widetilde{\Psi}\\
&-& [(i\partial_{\xi^{i}} - {\cal A}_{i})
g^{-\frac{1}{4}}\widetilde{\Psi}^{*}]\frac{\sqrt{g}g^{ij}}{2m}
[(-i\partial_{\xi^{j}} - {\cal A}_{j})
g^{-\frac{1}{4}}\widetilde{\Psi}]
\Big\}
\label{14}
\end{eqnarray}
An effective vector potential ${\cal A}_{i}(\bm\xi,t)$ in Eq.~(\ref{14}) is
defined as follows 
\begin{equation}
{\cal A}_{i}(\bm\xi,t) =  \widetilde{A}_{i}(\bm\xi,t)
+ \widetilde{v}_{i}(\bm\xi,t) - \partial_{\xi^{i}}S^{\text{cl}}(\bm\xi,t).
\label{15}
\end{equation}
The transformed Lagrangian, Eq.~(\ref{14}), can be also represented in
a common form of the Dirac-Frenkel functional. It is straightforward
to check that Eq.~(\ref{14}) is equivalent (up to irrelevant total
derivatives) to the functional
\begin{equation}
{\cal L} = \int\left[
i\widetilde{\Psi}^{*}\stackrel{\leftrightarrow}{\partial_{t}}
\widetilde{\Psi} 
- \widetilde{\Psi}^{*}\widetilde{H}[g_{ij},{\cal A}_{i}]
\widetilde{\Psi}\right] d{\bm\xi},
\label{16}
\end{equation}
where $\widetilde{H}[g_{ij},{\cal A}_{i}]$ is the Hamiltonian in the
moving frame: 
\begin{equation}
\widetilde{H} = g^{-\frac{1}{4}}
(-i\partial_{\xi^{i}} - {\cal A}_{i})\frac{\sqrt{g}g^{ij}}{2m}
(-i\partial_{\xi^{j}} - {\cal A}_{j})g^{-\frac{1}{4}}.
\label{17}
\end{equation}
Hence the transformed Schr\"odinger equation takes the form
\begin{equation} 
i\partial_{t}\widetilde{\Psi}(\bm\xi,t) 
= \widetilde{H}[g_{ij},{\cal A}_{i}]\widetilde{\Psi}(\bm\xi,t), 
\quad \widetilde{\Psi}(\bm\xi,0) = \Psi_{0}(\bm\xi).
\label{18}
\end{equation}

Equations (\ref{18}), (\ref{17}) and (\ref{15}) completely determine
the dynamics of a quantum particle in the local noninertial frame
moving with the velocity ${\bf v}({\bf x},t)$. The corresponding
Hamiltonian, Eq.~(\ref{17}), contains two types of ``external''
fields, a tensor field $g_{ij}$, and a vector field ${\cal
A}_{i}$. The tensor field $g_{ij}({\bm\xi},t)$ plays a role of an
effective metric, which produces a ``geodesic'' inertia force
\cite{TokPRB2005a}. This force makes a free particle to move along
geodesics in the deformed $\bm\xi$-space. The effective vector
potential ${\cal A}_{i}({\bf x},t)$ is responsible for a combined
action of the physical external forces and the rest of inertia forces
(i.~e. the linear acceleration force, and the generalized Coriolis and
centrifugal forces)\cite{TokPRB2005a}.

To complete the discussion of dynamics in a general noninertial frame,
we present two fundamental conservation laws that follow from the equation of
motion, Eq.~(\ref{18}) (for a detailed derivation see Appendix). These
are the continuity equation
\begin{equation} 
\partial_{t}\widetilde{n} + \partial_{\xi^{k}}\widetilde{j}^{k} = 0,
\label{19}
\end{equation}
and the local momentum balance equation
\begin{equation} 
\partial_{t}\widetilde{j}_{k} -  \widetilde{j}^{i}
(\partial_{\xi^{i}}{\cal A}_{k} - \partial_{\xi^{k}}{\cal A}_{i})
+ \widetilde{n}\partial_{t}{\cal A}_{k}
+ \sqrt{g}\nabla_{i}\widetilde{P}^{i}_{k} = 0.
\label{20}
\end{equation}
In equations (\ref{19}) and (\ref{20}) $\widetilde{n} =
|\widetilde{\Psi}|^{2}$ is the probability density, and
$\widetilde{j}^{i}$ and $\widetilde{P}^{i}_{k}$ are the current
density and the stress tensor, respectively. The operator
$\nabla_{i}$ in Eq.~(\ref{20}) stands for the covariant derivative in a
space with metric $g_{ij}$ (see, for example
Ref.~\onlinecite{DubrovinI}). In particular the stress force,
$\nabla_{i}\widetilde{P}^{i}_{k}$, in Eq.~(\ref{20}) is the covariant
divergence of a second rank tensor:
\begin{equation} 
\nabla_{i}\widetilde{P}^{i}_{k} = \frac{1}{\sqrt{g}}
\partial_{\xi^{i}}\sqrt{g}\widetilde{P}^{i}_{k} 
- \frac{1}{2}\widetilde{P}^{ij}\partial_{\xi^{k}}g_{ij}.
\label{21}
\end{equation}
In Appendix we show that in general the current density
and the stress tensor can be expressed in terms of
the functional derivatives of the Hamiltonian 
$\widetilde{H}[g_{ij},\bm{\mathcal{A}}]$ with respect to the vector
potential and the metric tensor, respectively:
\begin{eqnarray} 
\widetilde{j}^{k}(\bm\xi,t) &=& -\langle\widetilde{\Psi}|
\frac{\delta\widetilde{H}[g_{ij},
\bm{\mathcal{A}}]}{\delta{\cal A}_{k}(\bm\xi,t)}
|\widetilde{\Psi}\rangle,
\label{22a}\\
\widetilde{P}^{ij}(\bm\xi,t) &=& -\frac{2}{\sqrt{g}}\langle\widetilde{\Psi}|
\frac{\delta\widetilde{H}[g_{ij},
\bm{\mathcal{A}}]}{\delta g_{ij}(\bm\xi,t)}
|\widetilde{\Psi}\rangle
\label{22b}
\end{eqnarray}
In the case of a single quantum particle Eqs.~(\ref{22a}) and (\ref{22b})
lead to the following explicit representations
\begin{eqnarray}
\widetilde{j}_{k} &=& -\frac{i}{2m}
(\widetilde{\Psi}^{*}\partial_{\xi^{k}}\widetilde{\Psi}
- \widetilde{\Psi}\partial_{\xi^{k}}\widetilde{\Psi}^{*})
- \frac{\widetilde{n}}{m}{\cal A}_{k},
\label{23}\\
\widetilde{P}_{ij} &=& \frac{1}{2m}\Big[
(\hat{K}_{i}g^{-\frac{1}{4}}\widetilde{\Psi})^{*}
(\hat{K}_{j}g^{-\frac{1}{4}}\widetilde{\Psi}) + c.c.
\nonumber \\ 
&-& \frac{g_{ij}}{2\sqrt{g}}\partial_{\xi^{k}}\sqrt{g}g^{kl}
\partial_{\xi^{l}}\frac{\widetilde{\Psi}^{*}\widetilde{\Psi}}{\sqrt{g}}
\Big],
\label{24}
\end{eqnarray}
where $\hat{K}_{j} =-i\partial_{\xi^{j}}-{\cal A}_{j}$ is the
kinematic momentum.

\subsection{Quantum particle in a comoving frame}

The form of the equations of motion, Eqs.~(\ref{18}) and (\ref{15}), is
invariant under the transformation to an arbitrary local reference frame
defined by its velocity -- the vector-valued function ${\bf v}({\bf
  x},t)$. To specify a particular frame we need to supply the above
system of equations by a local ``gauge-fixing'' condition
\cite{TokPRB2005a}. One of the most simple and natural choices of such
a gauge condition is a requirement of zero current density:
\begin{equation}
\widetilde{\bf j}(\bm\xi,t) = 0.
\label{25}
\end{equation}
This requirement specifies a comoving Lagrangian frame. That is a
reference frame moving
with the velocity ${\bf v}({\bf x},t)={\bf j}({\bf x},t)/n({\bf
  x},t)$, where ${\bf j}$ and $n$ are the current and the density in
the laboratory frame. In this case the new coordinates $\bm\xi$ become
the Lagrangian coordinates, while the metric tensor $g_{ij}$,
Eq.~(\ref{7}), acquires a meaning of the Green's deformation tensor
\cite{PhysAc}. Substituting Eq.~(\ref{25}) in the 
the continuity equation we observe that in the Lagrangian frame the
density distribution is stationary and equals to the density
at $t=0$, which is fixed by the initial conditions
\begin{equation}
\widetilde{n}(\bm\xi,t) = \widetilde{n}(\bm\xi,0)
= n_{0}(\bm\xi) = |\Psi_{0}(\bm\xi)|^{2}.
\label{26}
\end{equation}
Similarly the local momentum balance equation, Eq.~(\ref{20}), in the
Lagrangian frame simplifies as follows
\begin{equation}
n_{0}\partial_{t}{\cal A}_{k}
+ \sqrt{g}\nabla_{i}\widetilde{P}^{i}_{k} = 0.
\label{27}
\end{equation}
Equation (\ref{27}) reveals a physical significance of the effective
vector potential ${\cal A}_{k}$. It produces a force (an effective
electric field) that exactly compensates the local stress force. As a
result the net force exerted on every infinitesimal volume element
vanishes, which guarantees vanishing current and a stationary density
in every point of the Lagrangian $\bm\xi$-space.

Using Eqs.~(\ref{23}) and (\ref{26}) we can represent the gauge condition of
Eq.~(\ref{25}) in the following explicit form
\begin{equation}
-\frac{i}{2m}(\widetilde{\Psi}^{*}\partial_{\xi^{k}}\widetilde{\Psi}
- \widetilde{\Psi}\partial_{\xi^{k}}\widetilde{\Psi}^{*})
= \frac{n_{0}}{m}{\cal A}_{k}.
\label{28}
\end{equation}

A complete set of equations of motion in the Lagrangian frame
consists of three equations. These are the Sch\"odinger equation
(\ref{18}), the zero-current 
condition (\ref{28}), and Eq.~(\ref{15}) that relates the
effective vector potential ${\cal A}$ to the external fields. The
solution of this system yields the wave function
$\widetilde{\Psi}(\bm\xi,t)$, and the trajectory, ${\bf x}(\bm\xi,t)$,
of the Lagrangian frame for a given configuration of the external
fields, ${\bf A}({\bf x},t)$ and $U({\bf x},t)$. 

The problem of finding ${\bf x}(\bm\xi,t)$ from Eq.~(\ref{15}) can be
brought to a more physical form. The time derivative of Eq.~(\ref{15})
takes the form
\begin{equation}
m\partial_{t}\widetilde{v}_{i} + \partial_{t}\widetilde{A}_{i}
- \partial_{\xi^{i}}
\left(\frac{m}{2}\widetilde{v}^{k}\widetilde{v}_{k} 
+ \widetilde{v}^{k}\widetilde{A}_{k} - U \right) 
- \partial_{t}{\cal A}_{i} = 0.
\label{29}
\end{equation}
The first three terms in the left hand side of Eq.~(\ref{29})
correspond to a combination of the external and the inertial forces,
while the last term is precisely equal to the local stress force [see
Eq.~(\ref{27})]. Hence  Eq.~(\ref{29}) [i.~e. the time derivative of
Eq.~(\ref{15})] has a clear meaning of the force balance equation in
the comoving frame. On the other hand, it  can be considered
as an equation of motion for the dynamic variable ${\bf
  x}(\bm\xi,t)$. Indeed, using Eq.~(\ref{5}), and the explicit
representations for $\widetilde{v}^{i}$, Eq.~(\ref{8}), and for
$\widetilde{A}_{i}$, Eq.~(\ref{10}), one can straightforwardly reduce
Eq.~(\ref{29}) to the following form
\begin{eqnarray} \nonumber
m\frac{\partial^{2} x^{k}}{\partial t^{2}} -
\frac{\partial x^{i}}{\partial t}\left(
\frac{\partial A_{k}}{\partial x^{i}} - 
\frac{\partial A_{i}}{\partial x^{k}}\right)
&+& \left(\frac{\partial A_{k}}{\partial t}\right)_{\bf x} \\
- \frac{\partial U}{\partial x^{k}} 
&-& \frac{\partial \xi^{i}}{\partial x^{k}}
\frac{\partial {\cal A}_{i}}{\partial t} = 0,
\label{30}
\end{eqnarray}
where $(\partial A_{k}/\partial t)_{\bf x}$ means the time derivative
at fixed ${\bf x}$. Equation (\ref{30}) is exactly the Newton equation
for a classical particle moving in the presence of the external
electro-magnetic force and the stress force [the last term in the left
hand side of Eq.~(\ref{30})]. 

Thus the complete system of equations,
which determines the quantum dynamics in the Lagrangian frame, can be
rewritten as follows
\begin{eqnarray}
&& i\partial_{t}\widetilde{\Psi}(\bm\xi,t) 
= \widetilde{H}[g_{ij},{\cal A}_{i}]\widetilde{\Psi}(\bm\xi,t), 
\label{31} \\
&& {\cal A}_{k} = 
-\frac{i}{2n_{0}} (\widetilde{\Psi}^{*}\partial_{\xi^{k}}\widetilde{\Psi}
- \widetilde{\Psi}\partial_{\xi^{k}}\widetilde{\Psi}^{*}),
\label{32} \\
&& m\ddot{x}^{k} = [\dot{\bf x}\times{\bf B}({\bf x},t)]_{k} 
+ E_{k}({\bf x},t) 
+ \frac{\partial \xi^{i}}{\partial x^{k}}
\partial_{t}{\cal A}_{i}, 
\label{33}
\end{eqnarray}
where ${\bf E}({\bf x},t)$ and ${\bf B}({\bf x},t)$
are the external electric and magnetic fields, which are defined in a
usual way:
\begin{eqnarray}
{\bf E}({\bf x},t) &=& - \partial_{t}{\bf A}({\bf x},t) 
- \partial_{\bf x}U({\bf x},t),
\label{34}\\
{\bf B}({\bf x},t) &=& \partial_{\bf x}\times{\bf A}({\bf x},t).
\label{35}
\end{eqnarray}
The system of Eqs.~(\ref{31})--(\ref{33}) should be solved with the
initial conditions
\begin{equation}
\widetilde{\Psi}(\bm\xi,0) = \Psi_{0}(\bm\xi), \quad
{\bf x}(\bm\xi,0) = \bm\xi, \quad 
\dot{\bf x}(\bm\xi,0) = {\bf v}_{0}(\bm\xi), 
\label{36}
\end{equation}
where ${\bf v}_{0}(\bm\xi)$ is the velocity distribution in the
initial state $\Psi_{0}(\bm\xi)$. The Hamiltonian
$\widetilde{H}[g_{ij},{\cal A}_{i}]$ is defined after Eq.~(\ref{17}),
and the metric (deformation) tensor $g_{ij}$ is connected to the
solution, ${\bf x}(\bm\xi,t)$, of Eq.~(\ref{33}) via Eq.~(\ref{7}).

Interestingly, the whole system of Eqs.~(\ref{31})--(\ref{33}) can be
obtained from a single variational functional of the following form
\begin{eqnarray}
\widetilde{\cal L}
[\widetilde{\Psi},\bm{\mathcal A},{\bf x}] 
&=& \int d{\bm\xi}\Big\{
i\widetilde{\Psi}^{*}\stackrel{\leftrightarrow}{\partial_{t}}
\widetilde{\Psi} 
- \widetilde{\Psi}^{*}\widetilde{H}[g_{ij},{\cal A}_{i}]
\widetilde{\Psi}
\nonumber\\
&+& n_{0}(\bm\xi)\Big[\frac{m}{2}(\dot{\bf x})^{2} 
+ \dot{\bf x}{\bf A}({\bf x},t) - U({\bf x},t)\Big]\Big\}
\label{37}
\end{eqnarray}
Apparently the first two conditions for the extremum of Eq.~(\ref{37}),
$\delta\widetilde{\cal L}/\delta\widetilde{\Psi}^{*}=0$ and
$\delta\widetilde{\cal L}/\delta\bm{\mathcal A}=0$, are equivalent 
to the Schr\"odinger equation, Eq.~(\ref{31}), and the zero-current
constraint, Eq.~(\ref{32}), respectively.  The third condition,
$\delta\widetilde{\cal L}/\delta{\bf x}=0$, yields the equation 
\begin{equation}
m\ddot{x}^{k} = [\dot{\bf x}\times{\bf B}({\bf x},t)]_{k} 
+ E_{k}({\bf x},t) - \frac{1}{n_{0}}\langle\widetilde{\Psi}|
\frac{\delta\widetilde{H}}{\delta x^{k}}|\widetilde{\Psi}\rangle. 
\label{38}
\end{equation}
By direct calculations one can check the following identity 
\begin{equation}
\langle\widetilde{\Psi}|
\frac{\delta\widetilde{H}}{\delta x^{k}}|\widetilde{\Psi}\rangle
= \frac{\partial \xi^{i}}{\partial x^{k}}
\sqrt{g}\nabla_{j}\widetilde{P}^{j}_{i}
= - \frac{\partial \xi^{i}}{\partial x^{k}}
n_{0}\partial_{t}{\cal A}_{i},
\label{39}
\end{equation}
where the variational definition of the stress tensor,
Eq.(\ref{22b}), and the local momentum balance equation of
Eq.~(\ref{27}) have been used. Hence the last term in the right hand
side of Eq.~(\ref{39}) is identical to the correct stress force
entering the equation of motion for ${\bf x}(\bm\xi,t)$,
Eq.~(\ref{33}).

The Lagrangian $\widetilde{\cal L}$, Eq.~(\ref{37}), describes
classical dynamics of infinitesimal fluid elements, coupled to
constrained quantum dynamics in a space with metric $g_{ij}$. It is
worth noting that the coupling is purely geometric (minimal) -- the
classical trajectories ${\bf x}(\bm\xi,t)$ enter the quantum part of
the problem only via the induced metric $g_{ij}(\bm\xi,t)$,
Eq.~(\ref{7}). 

The complete set of Eqs.~(\ref{31})--(\ref{33}) consists of two parts:
(i) the {\em universal}, i.~e. independent of external fields, quantum
problem defined by Eqs.~(\ref{31}), (\ref{32}), and (ii) the
``classical'' equation of motion, Eq.~(\ref{33}), for the trajectory
of the comoving frame. The universal problem of Eqs.~(\ref{31}),
(\ref{32}) corresponds to quantum dynamics in the space with a
given time-dependent metric, subjected to a constraint of zero current
density. The solution of this problem (provided it exists and unique)
determines the wave function and 
the selfconsistent vector potential as universal functionals of the metric
tensor, $\widetilde{\Psi}[g_{ij}](\bm\xi,t)$, and 
$\bm{\mathcal A}[g_{ij}](\bm\xi,t)$. Hence the stress force, the
stress tensor $\widetilde{P}_{ij}$, as well as any other observable
are also functionals of the metric (deformation) tensor. Substituting
the solution of the universal quantum problem into Eq.~(\ref{33}) we
obtain a closed equation of motion for the trajectory ${\bf
  x}(\bm\xi,t)$ 
\begin{equation}
m\ddot{x}^{k} = [\dot{\bf x}\times{\bf B}({\bf x},t)]_{k} 
+ E_{k}({\bf x},t) - 
\frac{\sqrt{g}}{n_{0}} \frac{\partial \xi^{i}}{\partial x^{k}}
\nabla_{j}\widetilde{P}^{j}_{i}[g_{ij}]
\label{40}
\end{equation}
Equation (\ref{40}) is easily recognized as an equation of a nonlinear
elasticity theory in the Lagrangian formulation of continuum mechanics
\cite{PhysAc}. In this context the functional dependence of the stress tensor
$\widetilde{P}^{j}_{i}$ on the deformation tensor $g_{ij}$ plays a
role of the exact equation of state. This equation of state is
determined from the solution of the universal quantum problem defined by
Eqs.~(\ref{31}), (\ref{32}). 

In the one particle case the universal problem of Eqs.~(\ref{31}),
(\ref{32}) is exactly solvable. Therefore the exact quantum equation
of state can be found in an explicit form. Indeed, it is easy
to see that the following wave function
and the selfconsistent vector potential
\begin{equation}
\widetilde{\Psi}(\bm\xi,t) = \sqrt{n_{0}(\bm\xi)}
e^{i\varphi(\bm\xi,t)}, \quad
{\mathcal A}_{k}(\bm\xi,t) = \partial_{\xi^{k}}\varphi(\bm\xi,t)
\label{41}
\end{equation}
satisfy the system of Eqs.~(\ref{31}), (\ref{32}) if the phase
$\varphi(\bm\xi,t)$ takes the form 
\begin{equation}
\varphi = \varphi_{0}(\bm\xi) +
\frac{1}{2m}\int_{0}^{t}\Big[g^{-\frac{1}{4}}\partial_{\xi^{i}}
\sqrt{g}g^{ij}\partial_{\xi^{j}}g^{-\frac{1}{4}}\sqrt{n_{0}}
\Big]dt'
\label{42}
\end{equation}
where $n_{0}(\bm\xi)$ and $\varphi_{0}(\bm\xi)$ are the density and
the phase of the initial state: $\Psi_{0}(\bm\xi) =
\sqrt{n_{0}}e^{i\varphi_{0}}$. Substituting the wave function,
Eq.~(\ref{41}), into Eq.~(\ref{24}) we find the stress tensor
functional (the quantum equation of state):
\begin{eqnarray}\nonumber
\widetilde{P}_{ij}[g_{ij}] &=& \frac{1}{m}\Big[
\big(\partial_{\xi^{i}}g^{-\frac{1}{4}}\sqrt{n_{0}}\big)
\big(\partial_{\xi^{j}}g^{-\frac{1}{4}}\sqrt{n_{0}}\big) \\
&-& \frac{g_{ij}}{4\sqrt{g}}\partial_{\xi^{k}}\sqrt{g}g^{kl}
\partial_{\xi^{l}}\frac{n_{0}}{\sqrt{g}}
\Big]
\label{43}
\end{eqnarray}
The corresponding stress force, which enters the ``elastic'' equation of
motion, Eq.~({\ref{40}}), can be calculated by taking either the
covariant divergence of the stress tensor, Eq.~(\ref{43}), or the time
derivative of the 
vector potential $\bm{\mathcal A}$, Eq.~(\ref{41}). The result takes
the form
\begin{equation}
\frac{\sqrt{g}}{n_{0}} 
\nabla_{j}\widetilde{P}^{j}_{k}  =
\frac{1}{2m}\partial_{\xi^{k}}\Big[g^{-\frac{1}{4}}\partial_{\xi^{i}}
\sqrt{g}g^{ij}\partial_{\xi^{j}}g^{-\frac{1}{4}}\sqrt{n_{0}}\Big]
\label{44}
\end{equation}
It is worth mentioning that the equation of motion, Eq.~(\ref{40}),
with the stress force of Eq.~(\ref{44}) can be
interpreted as the Lagrangian formulation of the one particle quantum
fluid dynamics (see, for example, Ref.~\onlinecite{Holland2005}). 

To complete the formal consideration of the one particle dynamics, we
note that Eq.~(\ref{40}) with the stress tensor of Eq.~(\ref{43})
corresponds to the Euler-Lagrange equation for the following
``elastic'' Lagrangian  
\begin{eqnarray} \nonumber
{\cal L}_{\text el}[{\bf x}] &=& \int d{\bm\xi}\Big\{
n_{0}\Big[\frac{m}{2}(\dot{\bf x})^{2}
+ \dot{\bf x}{\bf A}({\bf x},t) - U({\bf x},t)\Big] \\
&+& \big(\partial_{\xi^{i}}g^{-\frac{1}{4}}\sqrt{n_{0}}\big)
\frac{\sqrt{g}g^{ij}}{2m}
\big(\partial_{\xi^{j}}g^{-\frac{1}{4}}\sqrt{n_{0}}\big)
\Big\}
\label{45}    
\end{eqnarray}    
The last term in Eq.~(\ref{45}) plays a role of a quantum elastic
energy. As it should be, the elastic energy depends on 
${\bf x}(\bm\xi,t)$ only via the Green's deformation tensor
$g_{ij}(\bm\xi,t)$. 

\subsection{Overview of the main results and discussion}

Let us summarize main results of the present
section. Starting from the usual Dirac-Frenkel variational principle,
we derived a complete set of
Eqs.~(\ref{31})--(\ref{33}), which describes quantum dynamics
in the comoving Lagrangian frame. This
set of equations is generated by a generalized Dirac-Frenkel
functional, Eq.~(\ref{37}), that depends on three functions
$\widetilde{\Psi}(\bm\xi,t)$, ${\bf x}(\bm\xi,t)$, and $\bm{\mathcal
  A}(\bm\xi,t)$. Two of them, the wave function
$\widetilde{\Psi}(\bm\xi,t)$, and the frame's trajectory ${\bf
  x}(\bm\xi,t)$, enter the theory as dynamic variables, while the
effective vector potential $\bm{\mathcal A}(\bm\xi,t)$ is responsible
for the zero-current constraint. This constraint ensures that our
reference frame is indeed Lagrangian (comoving), i.~e. a special
frame where one observes no current, and a stationary density
distribution. In fact, the force produced by the vector potential
$\bm{\mathcal A}$ exactly compensates the stress force thus
providing a physical mechanism of the vanishing current density
[see Eq.~(\ref{27})]. 

The structure of the basic Lagrangian, Eq.~(\ref{37}), is extremely
simple and transparent. It describes the dynamics of two coupled
subsystems. These are: (i) the classical system of infinitesimal fluid
elements labeled by their initial positions $\bm\xi$, and
moving along trajectories ${\bf x}(\bm\xi,t)$; and (ii) the quantum
system placed in a space with metric $g_{ij}(\bm\xi,t)$, and subjected
to the zero-current constraint. The two systems are coupled via
Eq.~(\ref{7}) that identifies the metric $g_{ij}(\bm\xi,t)$
(entering the quantum problem) with the Green's
deformation tensor generated by the classical trajectories ${\bf
  x}(\bm\xi,t)$.  

One of the most important observations is that the constrained quantum
problem, Eqs.~(\ref{31}) and (\ref{32}), does not contain any physical
external potential, but depends only on the metric (deformation)
tensor. In that sense the quantum problem in the Lagrangian frame is
{\em universal} -- it defines the wave function as a {\em universal}
functional of the deformation tensor,
$\widetilde{\Psi}(\bm\xi,t)=\widetilde{\Psi}[g_{ij}](\bm\xi,t)$. As a
result any observable in the comoving frame is also a
universal functional of $g_{ij}$. It is, therefore, natural to call
this formalism a {\em time-dependent deformation functional theory}
(TDDefFT). The physical observable of primary importance
is the stress tensor $\widetilde{P}_{ij}$ as it determines
the stress force in the equation of motion for the fluid elements.
The existence of the universal functional $\widetilde{P}_{ij}[g_{ij}]$
[i.~e. the existence of the solution to the universal problem of
Eqs.~(\ref{31}), (\ref{32})] allows us to formulate a closed
hydrodynamics-type equation for the only collective variable, the
trajectory ${\bf x}(\bm\xi,t)$. It should be noted that the
knowledge of ${\bf x}(\bm\xi,t)$ is equivalent to the knowledge to the
time-dependent density $n({\bf x},t)$, and velocity ${\bf v}({\bf
  x},t)$ in the laboratory frame. The later quantities can be
recovered from the former one as follows
\begin{eqnarray}
n({\bf x},t) &=& \left[
\frac{n_{0}(\bm\xi)}{\sqrt{g(\bm\xi,t)}}
\right]_{\bm\xi = \bm\xi({\bf x},t)}, 
\label{46} \\ 
{\bf v}({\bf x},t) &=& \left[
\frac{\partial{\bf x}(\bm\xi,t)}{\partial t}
\right]_{\bm\xi = \bm\xi({\bf x},t)}. 
\label{47}     
\end{eqnarray}
where $\bm\xi({\bf x},t)$ is the inverse of ${\bf x}(\bm\xi,t)$. 

In the present section we considered the simplest, but still
nontrivial case of one particle quantum dynamics. An illustrative
power of this example is related to the possibility to exactly solve the
universal quantum problem in the Lagrangian frame, and to find the
universal functionals, $\widetilde{\Psi}[g_{ij}](\bm\xi,t)$ and
$\widetilde{P}_{ij}[g_{ij}](\bm\xi,t)$, in the explicit form [see
Eqs.~(\ref{41})-(\ref{42}) and Eq.~(\ref{43}), respectively]. The
exact solution of the universal problem is no longer possible if more
than one particle is present. However, as we will see in the next
section, both the general idea of TDDefFT and the formal structure of
the theory remain basically unchanged in the most general case of an
interacting quantum many-body system.

\section{Time-dependent deformation functional theory}

\subsection{Quantum many-body theory in the comoving frame}

Let us consider a system of $N$ identical particles interacting via a
two-body potential $V(|{\bf x}-{\bf x'}|)$. The system is described by
the $N$-body wave function $\Psi({\bf x}_{1},\dots,{\bf x}_{N},t)$ that
satisfies the time-dependent Schr\"odinger equation with the
Hamiltonian 
\begin{equation} 
H = H_{0} + H_{\text{int}},
\label{48}
\end{equation}
where $H_{0}$ corresponds to the sum of one particle contributions, and
$H_{\text{int}}$ is the interaction Hamiltonian:
\begin{eqnarray} 
H_{0} &=& \sum_{\alpha = 1}^{N}\left[\frac{(-i\partial_{\bf x_{\alpha}} 
- {\bf A}({\bf x_{\alpha}},t))^{2}}{2m}
+ U({\bf x}_{\alpha},t)\right],
\label{49} \\
H_{\text{int}} &=& \frac{1}{2}\sum_{\alpha,\beta}
V(|{\bf x}_{\alpha}-{\bf x}_{\beta}|). 
\label{50}
\end{eqnarray}
Following the route outlined in Sec.~II, we restate the problem in
a form of the variational principle with the following Lagrangian
\begin{equation}
{\cal L}[\Psi^{*},\Psi] = \int\left[
i\Psi^{*}\stackrel{\leftrightarrow}{\partial_{t}}\Psi 
- \Psi^{*}H\Psi\right]\prod_{\alpha = 1}^{N} d{\bf x}_{\alpha}.
\label{51}
\end{equation}
The next step is to make a transformation to the comoving Lagrangian
frame. Formally this corresponds to the transformation of coordinates
${\bf x}_{\alpha}\mapsto{\bm\xi}_{\alpha}$: 
${\bf x}_{\alpha} = {\bf x}({\bm\xi}_{\alpha},t), \alpha=1,\dots,N$,
where the function ${\bf x}({\bm\xi},t)$ is the trajectory of a fluid
element, which is defined by Eq.~(\ref{5}). In addition, we introduce
a renormalized many-body wave function
$\widetilde{\Psi}({\bm\xi}_{1},\dots,{\bm\xi}_{N},t)$ in the new
frame:
\begin{eqnarray}\nonumber
\Psi({\bf x}({\bm\xi}_{1},t),\dots,{\bf x}({\bm\xi}_{N},t),t) 
&=& \prod_{\alpha = 1}^{N}g^{-\frac{1}{4}}({\bm\xi}_{\alpha},t)
e^{iS_{\text{cl}}({\bm\xi}_{\alpha},t)}\\
&\times&\widetilde{\Psi}({\bm\xi}_{1},\dots,{\bm\xi}_{N},t),
\label{52}
\end{eqnarray}
where $iS_{\text{cl}}({\bm\xi},t)$ is the classical action defined
after Eq.~(\ref{13}). Equation~(\ref{52}) is a direct
generalization of Eq.~(\ref{12}) for the N-particle system. The rest
of calculations also straightforwardly follows the line of the
previous section. Namely, we substitute Eq.~(\ref{52}) into the
Lagrangian, Eq.~(\ref{51}), perform the abovementioned transformation
of coordinates, and successively repeat all intermediate steps
described in Sec.~II. As a result we arrive at the following
generalized Dirac-Frenkel functional $\widetilde{\cal L}
[\widetilde{\Psi}^{*},\widetilde{\Psi},\bm{\mathcal A},{\bf x}]$ that
describes the dynamics of $N$-particle system in the comoving frame,
\begin{eqnarray}
\widetilde{\cal L}
&=& \int\Big[
i\widetilde{\Psi}^{*}\stackrel{\leftrightarrow}{\partial_{t}}
\widetilde{\Psi} 
- \widetilde{\Psi}^{*}\widetilde{H}[g_{ij},\bm{\mathcal A}]
\widetilde{\Psi}\Big]\prod_{\alpha = 1}^{N} d{\bm\xi}_{\alpha}
\nonumber\\
&+& \int n_{0}(\bm\xi)\Big[\frac{m}{2}(\dot{\bf x})^{2} 
+ \dot{\bf x}{\bf A}({\bf x},t) - U({\bf x},t)\Big]d\bm\xi
\label{53}
\end{eqnarray}
where $n_{0}(\bm\xi)=N\int\prod_{\alpha = 2}^{N} d{\bm\xi}_{\alpha}
|\Psi_{0}({\bm\xi},{\bm\xi}_{2},\dots,{\bm\xi}_{N})|^{2}$
is the initial density distribution.

Equation~(\ref{53}) has precisely the same structure as the one
particle Lagrangian of
Eq.~(\ref{37}). The only difference is that the Hamiltonian
$\widetilde{H}[g_{ij},\bm{\mathcal A}]$ in Eq.~(\ref{53}) corresponds
to a system of $N$ interacting particles in the space with metric
$g_{ij}$. Accordingly $\widetilde{H}[g_{ij},\bm{\mathcal A}]$ contains
two terms,
\begin{equation} 
\widetilde{H}[g_{ij},\bm{\mathcal A}] =
\widetilde{H}_{0}[g_{ij},\bm{\mathcal A}] +
\widetilde{H}_{\text{int}}[g_{ij}]. 
\label{54}
\end{equation}
The first term, $\widetilde{H}_{0}$, is a sum of $N$ one particle
Hamiltonians:
\begin{equation}
\widetilde{H}_{0}[g_{ij},\bm{\mathcal A}] 
= \sum_{\alpha = 1}^{N}g^{-\frac{1}{4}}_{\alpha}
\hat{K}_{\alpha,i}\frac{\sqrt{g_{\alpha}}g^{ij}_{\alpha}}{2m}
\hat{K}_{\alpha,j}g^{-\frac{1}{4}}_{\alpha},
\label{55}
\end{equation}
where $\hat{K}_{\alpha,j}=-i\partial_{\xi^{j}_{\alpha}}
- {\cal A}_{j}(\bm\xi_{\alpha},t)$, and
$g^{ij}_{\alpha}=g^{ij}(\bm\xi_{\alpha},t)$.
The second term, $\widetilde{H}_{\text{int}}$, in Eq.~(\ref{54})
describes a pairwise interparticle interaction in the Lagrangian frame,
\begin{equation}
\widetilde{H}_{\text{int}}[g_{ij}] 
= \frac{1}{2}\sum_{\alpha,\beta}
V(l_{\bm\xi_{\alpha}\bm\xi_{\beta}}),
\label{56}
\end{equation}
where $l_{\bm\xi_{\alpha}\bm\xi_{\beta}}$ is the length of geodesic that
connects points $\bm\xi_{\alpha}$ and $\bm\xi_{\beta}$ in the space
with metrics $g_{ij}$. 

Apparently the physical significance of the Lagrangian,
Eq.~(\ref{53}), remains unchanged. It describes the classical system of
fluid elements, which is geometrically coupled the constrained
quantum $N$-body system in the space with metrics $g_{ij}$. The
Euler-Lagrange equations take the form
\begin{eqnarray}
&& i\partial_{t}\widetilde{\Psi}({\bm\xi}_{1},\dots,{\bm\xi}_{N},t)
= \widetilde{H}[g_{ij},{\cal A}_{i}]
\widetilde{\Psi}({\bm\xi}_{1},\dots,{\bm\xi}_{N},t) 
\label{57} \\
&& {\cal A}_{k}(\bm\xi,t) = -\frac{iN}{2n_{0}(\bm\xi)} 
\Im\int\prod_{\alpha = 2}^{N} d{\bm\xi}_{\alpha}
\nonumber \\
&&\qquad \times
\widetilde{\Psi}^{*}({\bm\xi},{\bm\xi}_{2},\dots,{\bm\xi}_{N},t)
\partial_{\xi^{k}}
\widetilde{\Psi}({\bm\xi},{\bm\xi}_{2},\dots,{\bm\xi}_{N},t)
\label{58} \\
&& m\ddot{x}^{k} = [\dot{\bf x}\times{\bf B}({\bf x},t)]_{k} 
+ E_{k}({\bf x},t) - 
\frac{\sqrt{g}}{n_{0}} \frac{\partial \xi^{i}}{\partial x^{k}}
\nabla_{j}\widetilde{P}^{j}_{i}
\label{59}
\end{eqnarray}
where $\widetilde{P}^{ij}(\bm\xi,t)$ is the expectation value of the stress
tensor operator [see Eq.~(\ref{22b})]. An explicit expression for the
stress tensor in the Lagrangian frame can be found in
\cite{TokPRB2005a,TokPRB2005b}. It is worth reminding that the stress force in
Eq.~(\ref{59}) can be represented in three equivalent forms,
Eq.~(\ref{39}).

The most important property of the set of Eqs.~(\ref{57})--(\ref{59})
is that the equations of constrained quantum dynamics, Eqs.~(\ref{57})
and (\ref{58}), constitute a closed {\em universal} problem,
which depends only on the initial state
$\Psi_{0}({\bm\xi}_{1},\dots,{\bm\xi}_{N})$, and on the
metric tensor $g_{ij}(\bm\xi,t)$. The solution to this problem,
provided it exists and unique, defines the
many body wave function as a {\em universal} functional of
$\Psi_{0}$ and $g_{ij}$,
i.~e. $\widetilde{\Psi}[\Psi_{0},g_{ij}]
({\bm\xi}_{1},\dots,{\bm\xi}_{N},t)$. The existence of this
functional is the key statement of TDDefFT. It implies the existence
of the exact nonequilibrium equation of state,
$\widetilde{P}^{ij}=\widetilde{P}^{ij}[g_{ij}]$, and thus allows to
formulate a closed theory of one vector-valued collective variable -- the
trajectory function ${\bf x}(\bm\xi,t)$. This theory can be
interpreted as the exact quantum continuum mechanics. 

\subsection{The universal problem and mapping 
theorems of TDDefFT}

The complete system of Eqs.~(\ref{57})-(\ref{59}) is simply a
reformulation of the original linear Schr\"odinger equation with the
Hamiltonian (\ref{48}). Therefore Eqs.~(\ref{57})-(\ref{59})
possess a unique solution, 
provided the external fields ${\bf A}$ and $U$, and the initial state
$\Psi_{0}$ are given. It is, however, not obvious that given
metric $g_{ij}$ we can solve the universal problem of Eqs.~(\ref{57}),
(\ref{58}) independently of the third equation,
Eq.~(\ref{59}). The problem of well-possedness of the nonlinear Cauchy
problem defined by Eqs.~(\ref{57}), (\ref{58}) for a given
$g_{ij}(\bm\xi,t)$ is equivalent to the problem of existence of
TDDefFT. 

In Sec.~IIB we have demonstrated that in the one particle case the
universal problem of Eqs.~(\ref{57}), (\ref{58}) [for $N=1$ they
reduce to Eqs.~(\ref{31}), (\ref{32})] admits an exact analytic
solution. Hence in this particular case the universal functional
$\widetilde{\Psi}[\Psi_{0},g_{ij}]$ (and thus TDDefFT) does
indeed exist. The simplest way to argue in favor TDDefFT in the
general case of $N>1$ is to refer to the Runge-Gross theorem
\cite{RunGro1984}, or to its generalizations \cite{Vignale2004}. In fact,
just this argumentation was used in the previous works on ``geometric''
TDDFT \cite{TokPRB2005b,TokLNP2006}. The present formulation of the
many-body theory opens up a possibility to approach the problem in an
alternative and, possibly, more natural and internally consistent way.

As we have already mentioned, within the present formalism the
question of existence of TDDefFT formally translates to the problem of
the existence and uniqueness of solutions to the universal quantum
problem, Eqs.~(\ref{57}), (\ref{58}).  This problem defines a map
$g_{ij}\mapsto\widetilde{\Psi}$, which also assumes a map
$g_{ij}\mapsto{\bm{\mathcal A}}$, and implies the existence of the
generalized equation of state. Substituting the constraint of
Eq.~(\ref{58}) into Eq.~(\ref{57}) we observe that the later becomes a
nonlinear Schr\"odinger equation (NSE) with a special type of cubic
nonlinearity. A rigorous analysis of the Cauchy problem for this NSE
will be presented in a separate publication \cite{TokNes}. In this
paper we adopt common in TDDFT field simplifying assumptions,
\cite{RunGro1984,vanLeeuwen1999,Vignale2004,TDDFT2006} and prove the
uniqueness of the solution to Eqs.~(\ref{57}), (\ref{58}) for analytic
and v-representable metrics $g_{ij}(\bm\xi,t)$.  As usual we call an
observable v-representable if it can be produced in a given physical
system by applying some external potentials. In the present context a
metric (deformation) tensor $g_{ij}$ is v-representable if it is
defined by Eqs.~(\ref{7}) and (\ref{5}), where ${\bf v}({\bf x},t)$ is
a physical velocity generated by some external potentials ${\bf
A}({\bf x},t)$ and $U({\bf x},t)$.

Let us assume that the deformation tensor is analytic in $t$. Hence it
possesses a Taylor expansion with a finite radius of convergence,
\begin{equation}
g_{ij}(t) = \delta_{ij} + \sum_{k=1}^{\infty}\frac{g_{ij}^{(k)}}{k!}t^{k},  
\label{60}
\end{equation}
where $g_{ij}^{(k)}=\partial_{t}^{k}g_{ij}(0)$. The first term in the
right hand side of Eq.~(\ref{60}) is the initial metric tensor 
$g_{ij}^{(0)}=g_{ij}(0)=\delta_{ij}$, which follows from the initial
condition to Eq.~(\ref{5}). Similarly one can expand the wave function
$\widetilde{\Psi}(t)$  
\begin{equation}
\widetilde{\Psi}(t) = \Psi_{0} 
+ \sum_{k=1}^{\infty}\frac{\widetilde{\Psi}^{(k)}}{k!}t^{k}  
\label{61}
\end{equation}
Now we substitute Eq.~(\ref{58}) into Eq.~(\ref{57}), and insert the
expansions Eqs.~(\ref{60}) and (\ref{61}) into the resulting nonlinear
Schr\"odinger equation. Collecting terms with the same power of $t$ we
transform 
Eq.~(\ref{58}) to the following set of equations for the derivatives
$\widetilde{\Psi}^{(k)}$
\begin{equation}
\widetilde{\Psi}^{(k)} = -i F^{(k-1)}, \qquad k=1,2,\dots  
\label{62}
\end{equation}
where $F^{(k)}$ denote the coefficients for the Taylor
expansion of the right hand side of Eq.~(\ref{57}). Since
Eq.~(\ref{58}), which relates $\bm{\mathcal{A}}$ to
$\widetilde{\Psi}$, is local in time, the nonlinear operator in the
right hand side of Eq.~(\ref{57}) also locally depends on the wave
function. This implies that the $k$th Taylor coefficient $F^{(k)}$ contains
derivatives $\widetilde{\Psi}^{(p)}$ only with $p\le k$. Hence
Eq.~(\ref{62}) can be schematically represented as follows
\begin{eqnarray} \nonumber
\widetilde{\Psi}^{(1)} &=& -i F^{(0)}[g_{ij}^{(0)};\Psi_{0}],
\\ \nonumber
\widetilde{\Psi}^{(2)} &=& -i F^{(1)}[g_{ij}^{(0)},g_{ij}^{(1)};
\Psi_{0},\widetilde{\Psi}^{(1)}],
\\ \nonumber
 &\cdots&
\\ \nonumber
\widetilde{\Psi}^{(k)} &=& -iF^{(k-1)}[g_{ij}^{(0)},\dots,g_{ij}^{(k-1)};
\Psi_{0},\dots,\widetilde{\Psi}^{(k-1)}],
\\ \nonumber
 &\cdots&
\end{eqnarray}   
Apparently this system can be solved recursively starting
from the first equation. The solution uniquely defines a map: 
$\{g_{ij}^{(k)},\Psi_{0}\}\mapsto\{\widetilde{\Psi}^{(k)}\}$.
Substituting the coefficients $\widetilde{\Psi}^{(k)}$ into the Taylor
expansion, Eq.~(\ref{61}), we obtain the time-dependent wave function as
a unique functional of the metric tensor and the initial state: 
$\widetilde{\Psi}[g_{ij},\Psi_{0}](t)$. Importantly, this
is true only if the Taylor series of Eq.~(\ref{61}) converges,
i~.e., if the solution to our nonlinear problem exists, which can not
be taken for granted in general. In this paper we follow the common
practice and make an additional assumption of v-representability
of the metric, which guarantees an {\em a priori} existence of the
solution and thus a convergence of the Taylor series. 

The above results constitute a constructive prove of following {\em
uniqueness theorem}: For an analytic and v-representable Green's
deformation tensor $g_{ij}$, the wave function $\widetilde{\Psi}(t)$
in the Lagrangian frame is a unique functional of $g_{ij}$ and the
initial state $\Psi_{0}$. In other words, the map
$g_{ij},\Psi_{0}\mapsto\widetilde{\Psi}(t)$ is unique.

This theorem is analogous to the Runge-Gross theorem in TDDFT. In
particular it proves the existence of the exact nonequilibrium
equation of state, $\widetilde{P}^{ij}=\widetilde{P}^{ij}[g_{ij}]$,
and the existence of a closed theory of only one collective variable
-- the trajectory ${\bf x}({\bm\xi,t})$. Since the deformation tensor
is a unique functional of the velocity, the above uniqueness theorem
can be also viewed as a proof of the velocity-to-wave function mapping
that forms a basis of TDCDFT.

The presented proof of $g_{ij}\mapsto\widetilde{\Psi}(t)$ mapping is
closely related to constructive proofs of the mapping theorems in
TDDFT and TDCDFT by van~Leeuwen \cite{vanLeeuwen1999} and Vignale
\cite{Vignale2004}, respectively. In fact, the formal statement of
the problem in Refs.~\onlinecite{vanLeeuwen1999} and
\onlinecite{Vignale2004} is very similar to our system of
Eqs.~(\ref{57}), (\ref{58}). In either proof one solves the many-body
Schr\"odinger equation, supplemented by a constraint -- the force
balance equation -- that relates the potential to a collective
variable of interest (the density in Ref.~\onlinecite{vanLeeuwen1999},
or the current in Ref.~\onlinecite{Vignale2004}). The structure of the
recursive calculation of the Taylor coefficients is basically the same
in all the proofs. However, in the present formulation of the theory
the proof becomes almost trivial due to a complete (both in time and
in space) locality of the constraint, Eq.~(\ref{58}). We note
that the proof of the Vignale's theorem can be also essentially
simplified by reformulating the problem in a similar local fashion
\cite{TokUnpub}. It 
is also important to note that neither proof attempts to address a question
of convergence of the resulting unique Taylor series for the potential
and/or the wave function. Therefore, strictly speaking, the
v-representability problem remains unresolved in any nonlinearized
version of TDDFT, in spite occasional statements to the contrary in
the literature.

\subsection{Keldysh-contour formulation of the exact quantum 
continuum mechanics} 

In the previous section we have proved that the wave function
$\widetilde{\Psi}({\bm\xi}_{1},\dots,{\bm\xi}_{N},t)$ and the
effective vector potential $\bm{\mathcal{A}}(\bm\xi,t)$ are universal
functionals of the Green's deformation tensor
$g_{ij}(\bm\xi,t)$. Hence the stress tensor $\widetilde{P}_{ij}$ in
the Lagrangian frame is also a universal functional of $g_{ij}$. In
general the stress tensor $\widetilde{P}_{ij}$ is proportional to the
expectation value of the {\em partial} variational derivative 
$\delta\widetilde{H}[g_{ij},\bm{\mathcal{A}}]/\delta g_{ij}$ at fixed
$\bm{\mathcal{A}}$ [see Eq.~(\ref{22b})]. However, in the Lagrangian
frame the current density vanishes, which implies the following identity
$j^{k}=\langle\delta\widetilde{H}[g_{ij},
\bm{\mathcal{A}}]/\mathcal{A}_{k}\rangle\equiv 0$. Thus in the
Lagrangian frame $\widetilde{P}_{ij}$
can be also defined via the {\em total} variational
derivative of the Hamiltonian with respect to the metric
\begin{equation}
\widetilde{P}^{ij}[g_{ij}](\bm\xi,t) = 
-\frac{2}{\sqrt{g}}\langle\widetilde{\Psi}[g_{ij}]|
\frac{\delta\widetilde{H}[g_{ij}]}{\delta g_{ij}(\bm\xi,t)}
|\widetilde{\Psi}[g_{ij}]\rangle,
\label{63}
\end{equation}
where $\widetilde{H}[g_{ij}]\equiv
\widetilde{H}[g_{ij},\bm{\mathcal{A}}[g_{ij}]]$. Equation (\ref{63})
relates the exact nonequilibrium equation of state to the
solution of the universal quantum problem, Eqs.~(\ref{57}),
(\ref{58}). Substituting this equation of state into Eq.~(\ref{59}) we
obtain a formally closed equation of the exact quantum
continuum mechanics (in the Lagrangian formalism)
\begin{equation}
m\ddot{x}^{k} = [\dot{\bf x}\times{\bf B}({\bf x},t)]_{k} 
+ E_{k}({\bf x},t) - 
\frac{\sqrt{g}}{n_{0}} \frac{\partial \xi^{i}}{\partial x^{k}}
\nabla_{j}\widetilde{P}^{j}_{i}[g_{ij}]
\label{64}
\end{equation}
By solving this equation with initial conditions ${\bf
x}(\bm\xi,0)=\bm\xi$ and $\dot{\bf x}(\bm\xi,0)={\bf v}_{0}(\bm\xi)$
we get a set of trajectories ${\bf x}(\bm\xi,t)$ for a given
configuration of the external fields. The knowledge of these
trajectories allows us to uniquely determine the density,
Eq.~(\ref{46}), and the velocity, Eq.~(\ref{47}), in the laboratory
frame.

The existence of a closed continuum mechanics defined by
Eq.~(\ref{64}) is a generic fact, which follows from the uniqueness
theorem of Sec.~IIIB. In spite of an apparent ``classical'' form of
Eq.~(\ref{64}), it exactly describes the dynamics of a quantum
many-body system. All quantum and correlation effects are encoded in
the equation of state -- the functional dependence of the stress
tensor on the deformation tensor.   

The question we address in this section concerns a principal possibility to
formulate the exact continuum mechanics in a form of a closed
variational principle. Namely, is there exists an action functional 
$S[{\bf x}(\bm\xi,t)]$ that generates the equation of motion,
Eq.~(\ref{63}). In Sec.~IIB we have explicitly constructed such a
functional for the exactly solvable one particle case. The
corresponding Lagrangian is given by Eq.~(\ref{45}) that defines a
simple, purely elastic, i.~e., local in time, theory. A formal reason
for this is the following identity
\begin{equation}
\langle\widetilde{\Psi}|
\frac{\delta\widetilde{H}[g_{ij}]}{\delta g_{ij}}
|\widetilde{\Psi}\rangle =
\frac{\delta}{\delta g_{ij}}\langle\widetilde{\Psi}|
\widetilde{H}[g_{ij}]|\widetilde{\Psi}\rangle 
\equiv \frac{\delta E[g_{ij}]}{\delta g_{ij}},
\label{65}
\end{equation}
where $E[g_{ij}]$ is the energy functional, which turns out to be
local in time [see the last term in Eq.~(\ref{45})]. Therefore in the
one particle case the stress tensor is equal to the variational
derivative of the energy functional, exactly as it is in the classical
elasticity theory. (We note in brackets that the space-nonlocality,
i.~e. the presence of gradients in
$E[g_{ij}]$, is responsible for quantum effects).

In a many/few-particle system the identity of Eq.~(\ref{65}) does not
hold. In general, for a system of $N>1$ particles the stress tensor is
not a functional derivative of any functional. Physically this is
related to a relative motion of particles, which produces a
non-instantaneous response of the system to a dynamic change of the
metric. Nonetheless a variational formulation of the theory is still
possible if one doubles the number of degrees of freedom by
considering the evolution along a Keldysh contour
\cite{Keldysh1965:e}.

The existence of the map $g_{ij},\Psi_{0}\mapsto\widetilde{\Psi}(t)$
assumes the existence of a unitary evolution operator $U[g_{ij}](t,0)$
\begin{eqnarray} 
|\widetilde{\Psi}(t)\rangle &=& U[g_{ij}](t,0)|\Psi_{0}\rangle,
\label{66}\\
U[g_{ij}](t,0) &=& 
T\exp\left\{-i\int_{0}^{t}\widetilde{H}[g_{ij}](t')dt'\right\},
\label{67}
\end{eqnarray}
where $T$ stands for the usual chronological ordering. Using
Eqs.~(\ref{66}) and (\ref{67}) we can rewrite the definition of the
stress tensor, Eq.~(\ref{63}), as follows
\begin{equation}
\widetilde{P}^{ij}= 
-\frac{2}{\sqrt{g}}\langle\Psi_{0}|U(0,t)
\frac{\delta\widetilde{H}[g_{ij}]}{\delta g_{ij}}
U(t,0)|\Psi_{0}\rangle.
\label{68}
\end{equation}
Let us introduce two different deformation tensors, $g^{-}_{ij}(t)$,
and $g^{+}_{ij}(t)$, and construct the following generating functional
\begin{equation}
W[g^{-}_{ij},g^{+}_{ij}] = 
i\ln\langle\Psi_{0}|U^{+}(0,\infty)U^{-}(\infty,0)|\Psi_{0}\rangle,
\label{69}
\end{equation}
where $U^{\pm}(t,0)=U[g^{\pm}_{ij}](t,0)$ is the evolution operator
obtained from the solution of the universal problem, Eqs.~(\ref{57}),
(\ref{58}), with the metric $g_{ij}^{\pm}(\bm\xi,t)$. The stress
tensor, Eq.~(\ref{68}), is recovered by differentiating
$W[g^{-}_{ij},g^{+}_{ij}]$ with respect to $g_{ij}^{-}$, and setting
$g_{ij}^{-}=g_{ij}^{+}=g_{ij}$. Formally the operator
$U^{+}(0,\infty)U^{-}(\infty,0)$ in Eq.~(\ref{69}) describes a
propagation from the initial time to infinity, and then back to
$t=0$. This can be viewed as a propagation along a closed Keldysh
contour $C$ \cite{Keldysh1965:e}. The contour $C$ consists of two
branches: the ``forward'' 
$(-)$ branch that goes from $0$ to $\infty$, and the ``back'' $(+)$
branch going from $\infty$ to the initial time, $t=0$. Using this
notion one can represent the generating functional $W$,
Eq.~(\ref{69}), in the following compact form
\begin{equation}
W[g^{C}_{ij}] = 
i\ln\langle\Psi_{0}|
T_{C}e^{-i\int_{C}\widetilde{H}[g_{ij}^{C}](t)dt}
|\Psi_{0}\rangle,
\label{70}
\end{equation}
where $T_{C}$ orders times along the Keldysh contour, and $g_{ij}^{C}$
takes the values $g_{ij}^{-}$, and  $g_{ij}^{+}$ on the forward, and
back branches, respectively. The physical stress tensor,
Eq.~(\ref{68}), is given by the following functional derivative
\begin{equation}
\widetilde{P}^{ij}[g_{ij}] = -\frac{2}{\sqrt{g}}\left.
\frac{\delta W[g^{C}_{ij}]}{\delta g^{C}_{ij}}
\right|_{g^{C}_{ij}=g_{ij}(t)}.  
\label{71}
\end{equation}
The notation $g^{C}_{ij}=g_{ij}(t)$ means that we set metric tensors on
either branch equal to the physical deformation tensor
$g_{ij}(\bm\xi,t)$. Let us introduce the contour trajectory 
${\bf x}_{C}(\bm\xi,t)$ that generates the contour deformation tensor
\begin{equation} 
g_{ij}^{C} = \frac{\partial x^{k}_{C}}{\partial\xi^{i}}
\frac{\partial x^{k}_{C}}{\partial\xi^{j}}.
\label{72}
\end{equation}
Using Eqs.~(\ref{71}) and (\ref{72}) one can show that
the stress force entering Eq.~(\ref{64}) is the functional
derivative of the generating functional $W$ with respect to ${\bf x}_{C}$
\begin{equation}
- n_{0}\widetilde{F}_{k}^{\text{str}} 
= \sqrt{g} \frac{\partial \xi^{i}}{\partial x^{k}}
\nabla_{j}\widetilde{P}^{j}_{i}[g_{ij}] = \left.
\frac{\delta W[g^{C}_{ij}]}{\delta x^{k}_{C}}
\right|_{{\bf x}_{C}={\bf x}(\bm\xi,t)}
\label{73}  
\end{equation}
Equation (\ref{73}) naturally suggests the following form of the
Keldysh action functional
\begin{eqnarray} \nonumber
S_{C}[{\bf x}_{C}] = \int_{C}dt\int d{\bm\xi}
n_{0}\Big[&&\frac{m}{2}(\dot{\bf x}_{C})^{2}
+ \dot{\bf x}_{C}{\bf A}({\bf x}_{C},t)\\
&& - U({\bf x}_{C},t)\Big] - W[g^{C}_{ij}].
\label{74}    
\end{eqnarray}    
Indeed, the stationarity condition for the action $S_{C}[{\bf x}_{C}]$,
\begin{equation}
\left.\frac{\delta S_{C}[{\bf x}_{C}]}{\delta {\bf x}_{C}}
\right|_{{\bf x}_{C}={\bf x}(\bm\xi,t)} = 0,
\label{75}
\end{equation}
recovers the correct form of the hydrodynamics equation of motion,
Eq.~(\ref{64}), for the trajectory ${\bf x}(\bm\xi,t)$. 

The variational formulation, Eqs.~(\ref{74}), (\ref{75}), of the exact
quantum continuum mechanics is the main result of the present
section. From the practical point of view, the very existence of the
action functional, Eq.~(\ref{74}), is already a very useful
statement. In particular it justifies an application of many powerful
methods of the classical continuum mechanics to dynamics of
quantum many-body systems. One of those methods is a
construction of an effective elastic functionals $W[g_{ij}]$ based on
fundamental symmetries of a given physical system. Recently this
approach has been used to derive a hydrodynamics theory of strongly
correlated many-body states in the fractional quantum Hall regime
\cite{TokPRB2006b}. One of the basic assumptions made in
Ref.~\onlinecite{TokPRB2006b} was the existence of the action functional
of the form of Eq.~(\ref{74}). The results of the present section
provide a rigorous justification of that approach. In the next section
we will show the variational formulation of TDDefFT also offers a
convenient tool for the definition of xc potentials in the Kohn-Sham
scheme. 

A few years ago a Keldysh-contour formulation of TDDFT was proposed by
van~Leeuwen \cite{vanLeeuwen1998,vanLeeuwen2001} to resolve the
causality problem of the original Runge-Gross theory. Despite certain
similarities, our construction is fundamentally different from that of
Refs.~\onlinecite{vanLeeuwen1998,vanLeeuwen2001}. The van~Leeuwen's
formulation of TDDFT requires the existence of the Keldysh-contour
analog of the Runge-Gross mapping theorem, which has not been proved up to
now. In contrast to that, the present approach to TDDefFT relies only
on the real-time uniqueness theorem presented in Sec.~IIIB.

\section{A time-dependent Kohn-Sham construction}

In general the exact stress tensor $\widetilde{P}_{ij}[g_{ij}]$ as
well as the effective energy functional $W[g_{ij}^{C}]$ contain both
kinetic and interaction contributions. In some systems, such as
strongly correlated collective quantum Hall states, or one-dimensional
Luttinger liquids, it is natural to consider the functional
$\widetilde{P}_{ij}[g_{ij}]$ (or $W[g_{ij}^{C}]$) as a single
entity. This approach was successfully employed in our recent studies
of the fractional quantum Hall liquids and liquid crystals
\cite{TokPRB2006a,TokPRB2006b,TokVig2006cond-mat}. However, in
the most of less exotic many-body systems, e.~g., in atoms, molecules or
solids, it is useful to extract at least a part of the kinetic
contribution to the universal functionals, and to consider it
separately from the rest. The Kohn-Sham (KS) construction is a special tool
for such a separation -- it allows one to calculate exactly the
noninteracting part of the kinetic stress functionals. 

The time-dependent KS construction in TDDefFT can be introduced as
follows. Let us consider a system of $N$ noninteracting KS particles
moving in the presence of effective potentials, 
${\bf A}_{\text{S}}={\bf A}+{\bf A}_{\text{xc}}$ and $U_{\text{S}} = U
+ U_{\text{xc}}$. Here ${\bf A}({\bf x},t)$ and $U({\bf x},t)$ are the
external fields, while ${\bf A}_{\text{xc}}({\bf x},t)$ and
$U_{\text{xc}}({\bf x},t)$ are selfconsistent xc potentials that are
adjusted to reproduce a collective variable of interest in the physical
interacting system. In the Lagrangian formulation of TDDefFT the
proper collective variable is the trajectory ${\bf x}(\bm\xi,t)$.
In the KS system the equation of motion for
${\bf x}(\bm\xi,t)$ takes the form
\begin{equation}
m\ddot{\bf x} = \dot{\bf x}\times{\bf B}_{\text{S}}({\bf x},t) 
+ {\bf E}_{\text{S}}({\bf x},t) 
+ \widetilde{\bf F}_{\text{S}}^{\text{str}}[{\bf x}],
\label{76}
\end{equation}
where $\widetilde{\bf F}_{\text{S}}^{\text{str}}[{\bf x}]$ is the
kinetic stress force that is related to the kinetic stress tensor,
$\widetilde{P}_{\text{S},ij}[g_{ij}]$, of noninteracting KS particles:
\begin{equation}
\widetilde{F}_{\text{S},k}^{\text{str}}[{\bf x}] 
= - \frac{\sqrt{g}}{n_{0}} \frac{\partial \xi^{i}}{\partial x^{k}}
\nabla_{j}\widetilde{P}^{j}_{\text{S}i}[g_{ij}]. 
\label{77}  
\end{equation}
In Eq.~(\ref{76}) ${\bf E}_{\text{S}}({\bf x},t)$, and ${\bf
  B}_{\text{S}}({\bf x},t)$ are, respectively, the electric, and the
magnetic fields associated to the effective potentials  ${\bf
  A}_{\text{S}}({\bf x},t)$ and $U_{\text{S}}({\bf x},t)$. The
functional $\widetilde{P}_{\text{S},ij}[g_{ij}]$ is obtained from the
solution of the universal problem, Eqs.~(\ref{57}), (\ref{58}), for a
noninteracting system (i.~e., with
$\widetilde{H}_{\text{int}}[g_{ij}]=0$). To determine the xc
potentials one has to compare Eq.~(\ref{76}) with
the corresponding equation for the real interacting system,
Eq.~(\ref{64}). Apparently they coincide if the force produced by the
xc potentials equals to the difference of stress forces in the
interacting and the noninteracting systems
\begin{equation}
  \label{78}
 E_{\text{xc},k}({\bf x},t)
+ [\dot{\bf x}\times{\bf B}_{\text{xc}}({\bf x},t)]_{k} 
= - \frac{\sqrt{g}}{n_{0}} \frac{\partial \xi^{i}}{\partial x^{k}}
\nabla_{j}\widetilde{P}^{j}_{\text{xc},i}[g_{ij}],  
\end{equation}
where $\widetilde{P}^{j}_{\text{xc},i}[g_{ij}](\bm\xi,t) =
\widetilde{P}^{j}_{i}[g_{ij}](\bm\xi,t) -
\widetilde{P}^{j}_{\text{S},i}[g_{ij}](\bm\xi,t)$ is the xc stress
tensor functional, and the xc electric and magnetic fields are defined
as follows
\begin{eqnarray}
{\bf E}_{\text{xc}}({\bf x},t) &=& 
- \partial_{t}{\bf A}_{\text{xc}}({\bf x},t) 
- \partial_{\bf x}U_{\text{xc}}({\bf x},t),
\label{79}\\
{\bf B}_{\text{xc}}({\bf x},t) &=& 
\partial_{\bf x}\times{\bf A}_{\text{xc}}({\bf x},t).
\label{80}
\end{eqnarray}
Equations (\ref{78})--(\ref{80}) define the xc potentials up to a
gauge transformation. 

In practical application it is much more convenient to work with the
KS system (i.~e., to solve the time-dependent KS equations) in the
laboratory frame. Therefore we need to transform the definition of xc
potentials from the Lagrangian frame back to the laboratory one. This
is done simply by setting $\bm\xi = \bm\xi({\bf x},t)$, where
$\bm\xi({\bf x},t)$ is the inverse of ${\bf x}(\bm\xi,t)$. The result
of this procedure for Eq.~(\ref{78}) takes the form
\begin{equation}
  \label{81}
\partial_{t}A_{\text{xc},k} -
({\bf v}\times(\partial_{\bf x}\times{\bf A}_{\text{xc}}))_{k} 
+ \partial_{k}U_{\text{xc}}
= \frac{1}{n}\partial_{j}P_{\text{xc},jk}  
\end{equation}
where $P_{\text{xc},ij}({\bf x},t)$ is the xc stress tensor in the
laboratory frame, which is related to
$\widetilde{P}_{\text{xc},ij}[g_{ij}](\bm\xi,t)$ as follows
\begin{equation}
  \label{82}
P_{\text{xc},ij}({\bf x},t) = \frac{\partial \xi^{k}}{\partial x^{i}} 
\frac{\partial \xi^{l}}{\partial x^{j}}
\widetilde{P}_{kl}[g_{ij}](\bm\xi({\bf x},t),t). 
\end{equation}

Equation (\ref{81}) recovers the force definition of the xc potentials
introduced in Ref.~\onlinecite{TokPRB2005a}. The most important new
result of the present general approach is the functional dependence on
the collective variables. We 
have proved that the stress tensor in the Lagrangian frame,
$\widetilde{P}_{\text{xc},ij}(\bm\xi,t)$,  is a unique
functional of only one basic variable -- the Green's deformation
tensor $g_{ij}$. The transformed xc stress tensor
$P_{\text{xc},ij}({\bf x},t)$, Eq.~(\ref{82}), which determines xc
potentials in the laboratory frame, already depends not only on
$g_{ij}$, but also on the function $\bm\xi({\bf x},t)$
itself. However, the dependence on $\bm\xi({\bf x},t)$ is trivial
[in the prefactor, and in the argument in Eq.~(\ref{82})], and can be
accounted for exactly, provided the universal functional
$\widetilde{P}_{\text{xc},ij}[g_{ij}](\bm\xi,t)$ is known. 

Another practically important outcome is the possibility to define the
xc force, and thus the xc potentials, via a functional derivative of
the scalar functional $W[g_{ij}^{\text{C}}]$. Indeed, using
Eq.~(\ref{73}) we can rewrite Eq.~(\ref{78}) in the following form
\begin{equation}
  \label{83}
 {\bf E}_{\text{xc}}
+ \dot{\bf x}\times{\bf B}_{\text{xc}}
= - \frac{1}{n_{0}} \left.
\frac{\delta W_{\text{xc}}[g^{C}_{ij}]}{\delta {\bf x}_{C}}
\right|_{{\bf x}_{C}={\bf x}(\bm\xi,t)}  
\end{equation}
where $W_{\text{xc}}[g^{C}_{ij}] = W[g^{C}_{ij}] -
W_{\text{S}}[g^{C}_{ij}]$ is the difference of W-functionals in the
interacting, and noninteracting systems. The variational definition of
the xc potentials, Eq.~(\ref{83}), should be more convenient for
phenomenological construction (e.~g. GGA-like) of various
approximations. It reduces the problem to approximating a global
scalar functional $W_{\text{xc}}[g^{C}_{ij}]$, which seems to be a
simpler task. Importantly, the very form of Eq.~(\ref{83}) already
guarantees many exact constraints. For instance, any functional
$W_{\text{xc}}[g^{C}_{ij}]$ yields the xc force that is equal to the
divergence of a symmetric second rank tensor, which automatically
ensures the zero net force, and the zero net torque conditions. It is
worth outlining that the variational definition of the xc potentials,
Eq.~(\ref{83}), is possible only in the Lagrangian frame. Given a
functional $W_{\text{xc}}[g^{C}_{ij}]$, the xc force in the laboratory
frame is obtained as follows. One first calculates the variational
derivative in the right hand side of Eq.~(\ref{83}), and then makes
the transformation to the laboratory frame by setting
$\bm\xi=\bm\xi({\bf x},t)$.  As a result the right hand side of
Eq.~(\ref{82}) is recovered.

The W-functional of TDDefFT also offers a convenient tool for a
compact, unified, and transparent representation of all currently known
approximations in TDDFT/TDCDFT. For example, the local VK approximation
\cite{VigKohn1996}, as well as its extension by VUC
\cite{VigUllCon1997}, corresponds to the following quadratic functional
\begin{eqnarray}
\nonumber
&&W_{\text{xc}}^{\text{VK}}[g^{C}_{ij}] = \frac{1}{8}\int d\bm\xi\int_{C}dtdt'
\Big\{2\mu_{\text{xc}}^{\text{C}}(t-t')
\delta g^{C}_{ij}(t)\delta g^{C}_{ij}(t')
\\
&& \qquad + \Big[K_{\text{xc}}^{\text{C}}(t-t')
- \frac{2}{3}\mu_{\text{xc}}^{\text{C}}(t-t')\Big]
\delta g^{C}_{ii}(t)\delta g^{C}_{jj}(t')  
\Big\}, 
\label{84}
\end{eqnarray}
where $\delta g_{ij} = g_{ij} - \delta_{ij}$ is the linearized strain
tensor, while $K_{\text{xc}}^{\text{C}}(t-t')$ and
$\mu_{\text{xc}}^{\text{C}}(t-t')$ are, respectively, the
nonadiabatic xc bulk and shear moduli of the homogeneous system,
defined on the Keldysh contour. The VK approximation given by
Eq.~(\ref{84}) is valid in the limit of small $\delta g_{ij}$.

A nonlinear elastic local deformation approximation (LDefA) introduced
in Ref.~\onlinecite{TokPRB2005b} is generated by a completely local
W-functional of the following form
\begin{equation}
W_{\text{xc}}^{\text{LDefA}}[g^{C}_{ij}] = \int d\bm\xi\int_{C}dt 
E_{\text{xc}}(g^{C}_{ij}(\bm\xi,t)), 
\label{85}
\end{equation}
where the nonadiabatic xc energy density
$E_{\text{xc}}(g_{ij})$ is defined as follows  
\begin{eqnarray}\nonumber
E_{\text{xc}}(g_{ij}) &=& \sum_{\bf p}\Big\{
g^{ij}\frac{p_{i}p_{j}}{2m}f_{\text{xc}}(p;n_{0}) \\
&& \qquad + \frac{1}{2\sqrt{g}} 
\bar{V}\left(\sqrt{g^{ij}p_{i}p_{j}}\right)g_{2}(p;n_{0})
\Big\}
\label{86}
\end{eqnarray}
In this equation $\bar{V}(q)$ is the Fourier component of the
interaction potential, $f_{\text{xc}}(p;n_{0})$ is the correlation
part of the one particle distribution function, and $g_{2}(p;n_{0})$
is the Fourier component of the pair correlation function. Both
$f_{\text{xc}}(p;n_{0})$ and $g_{2}(p;n_{0})$ are calculated for a
homogeneous system with the density $n_{0}(\bm\xi)$. In the limit of
small deformations, when $g_{ij}$ slightly deviates from
$\delta_{ij}$, the nonlinear elastic approximation defined by
Eqs.~(\ref{85}), (\ref{86}) reduces to the high-frequency limit of VK
approximation, Eq.~(\ref{84}).

For comparison we also show the functional
$W_{\text{xc}}^{\text{Ad}}[g^{C}_{ij}]$ that corresponds to the
adiabatic local density approximation (ALDA): 
\begin{equation}
W_{\text{xc}}^{\text{Ad}}[g_{ij}] = \int d\bm\xi\int dt 
\sqrt{g(\bm\xi,t)}E^{\text{hom}}_{\text{xc}}\left(
\frac{n_{0}(\bm\xi)}{\sqrt{g(\bm\xi,t)}}\right), 
\label{87}
\end{equation}
where $E^{\text{hom}}_{\text{xc}}(n)$ is the usual ground state xc energy
density of the homogeneous system. 

Finally we derive one more exact representation of the xc
force. Namely, we relate ${\bf A}_{\text{xc}}$ and $U_{\text{xc}}$ to
the effective vector potential $\bm{\mathcal A}$ that enters the
universal problem of Eqs.~(\ref{57}), (\ref{58}). According to the
momentum balance equation in the Lagrangian frame, Eq.~(\ref{27}),
[see also the identity of Eq.~(\ref{39})], the divergence of the
stress tensor is equal to the time derivative of the effective vector
potential $\bm{\mathcal A}$. Hence the xc force in the Lagrangian
frame, Eq.~(\ref{78}), can be also represented as follows
\begin{equation}
  \label{88}
 E_{\text{xc},k}({\bf x},t)
+ [\dot{\bf x}\times{\bf B}_{\text{xc}}({\bf x},t)]_{k} 
=  \frac{\partial \xi^{i}}{\partial x^{k}}
\partial_{t}{\mathcal A}_{\text{xc},i}(\bm\xi,t),  
\end{equation}
where $\bm{\mathcal A}_{\text{xc}}(\bm\xi,t) = \bm{\mathcal
  A}(\bm\xi,t) - \bm{\mathcal A}_{\text{S}}(\bm\xi,t)$  is the
difference of the effective vector potentials 
in the interacting and noninteracting systems with the same metric
$g_{ij}$. The right hand side of Eq.~(\ref{88}) needs to be
transformed to the laboratory frame. Let us first use the standard
transformation rule to define an effective
vector potential, $\bm{\mathcal A}'_{\text{xc}}({\bf x},t)$, in the
laboratory frame
\begin{equation}
{\mathcal A}_{\text{xc},i}(\bm\xi,t) = 
\frac{\partial x^{j}}{\partial \xi^{i}}
{\mathcal A}'_{\text{xc},j}({\bf x}(\bm\xi,t),t).
\label{89}
\end{equation} 
Substituting Eq.~(\ref{89}) in the the right hand side of
Eq.~(\ref{88}) we find for the stress force
\begin{eqnarray} \nonumber
&&\frac{\partial \xi^{i}}{\partial x^{k}}
\partial_{t}{\mathcal A}_{\text{xc},i}(\bm\xi,t) =
\left .\partial_{t}{\mathcal A}'_{\text{xc},k}\right|_{\bm\xi}
+ \frac{\partial \xi^{i}}{\partial x^{k}}
\frac{\partial^{2} {x}^{j}}{\partial t\partial \xi^{i}}
{\mathcal A}'_{\text{xc},j}\\ 
&& \qquad =\left .\partial_{t}{\mathcal A}'_{\text{xc},k}\right|_{\bf x}
+ v^{j}\frac{\partial {\mathcal A}'_{\text{xc},k}}{\partial x^{j}} +
\frac{\partial v^{j}}{\partial x^{k}}{\mathcal A}'_{\text{xc},j}
\label{90}
\end{eqnarray}
where we used the definition of the velocity, Eq.~(\ref{5}). Inserting
the result of Eq.~(\ref{90}) into Eq.~(\ref{88}) we obtain the
following equation for the xc potentials in the laboratory frame
\begin{eqnarray} \nonumber
\partial_{t}{\bf A}_{\text{xc}} &-&
{\bf v}\times(\partial_{\bf x}\times{\bf A}_{\text{xc}}) 
+ \partial_{\bf x}U_{\text{xc}}
= - \partial_{t}\bm{\mathcal A}'_{\text{xc}}\\ 
&+& {\bf v}\times(\partial_{\bf x}\times\bm{\mathcal A}'_{\text{xc}}) 
- \partial_{\bf x}({\bf v}\bm{\mathcal A}'_{\text{xc}}).
\label{91}
\end{eqnarray}
This equation determines the xc potentials, 
${\bf A}_{\text{xc}}({\bf x},t)$ and $U_{\text{xc}}({\bf x},t)$, up to
an arbitrary gauge transformation. One of possible solutions to
Eq.~(\ref{91}) takes the form
\begin{eqnarray}
{\bf A}_{\text{xc}}({\bf x},t) &=& 
- \bm{\mathcal A}'_{\text{xc}}({\bf x},t),
\label{92}\\
U_{\text{xc}}({\bf x},t) &=& - {\bf v}({\bf x},t)
\bm{\mathcal A}'_{\text{xc}}({\bf x},t).
\label{93}
\end{eqnarray}
Therefore there is a particular mixed gauge in which the xc potentials
are locally expressed in terms of the effective vector potential
$\bm{\mathcal A}_{\text{xc}}$. The convenience of the exact
representation given by Eqs.~(\ref{92}), (\ref{93}) is that it
directly relates the xc potentials, which enter the KS equations, to
the solution of the universal quantum problem, Eqs.~(\ref{57}),
(\ref{58}). Equations (\ref{92}) and (\ref{93}) can be useful for the
analysis of the exact properties of KS potentials, such as symmetries,
scaling properties, etc.

\section{Conclusion}

In this paper we presented a selfcontained, constructive derivation of
the time-dependent deformation functional theory (TDDefFT). The main
idea of our approach to the time-dependent many-body problem is a
separation of the convective and relative motions of quantum
particles. Technically these two types of motion are separated by the
transformation to the comoving Lagrangian reference frame. The
convective motion is described by a set of trajectories ${\bf
x}(\bm\xi,t)$ of infinitesimal fluid elements, where $\bm\xi$ is the
initial position (the Lagrangian coordinate) of a given
element. The motion of particles relatively to the convective flow is
determined by the many-body wave function $\widetilde{\Psi}$ in the
comoving frame. Since the convective motion is singled out by the
above transformation, the number of degrees of freedom entering the
quantum many-body problem is reduced. Formally the dynamics of the
wave function $\widetilde{\Psi}$ is constrained by a local ``gauge''
condition of zero current density. The most important property of this
constrained quantum problem is that it does not contain external
fields. It is completely determined by the fundamental geometric
characteristics of the Lagrangian frame -- the Green's deformation
tensor $g_{ij}$ that enters the equations of motion as a metric
tensor. Hence the many-body problem for the relative motion
appears to be universal. This problem, naturally defines the wave
function as a universal functional of the deformation tensor,
$\widetilde{\Psi}[g_{ij}]$. Therefore the expectation value of any
observable in the Lagrangian frame is also a functional of the
deformation tensor. In particular this is true for the stress force
entering the equation of motion for the Lagrangian trajectories ${\bf
x}(\bm\xi,t)$. Thus the trajectories and hence the whole convective
motion of an arbitrary quantum many-body system can be found from a
closed hydrodynamics-like theory. We call this theory TDDefFT since
the deformation tensor is the basic variable entering all relevant
universal functionals.

The set of Lagrangian trajectories ${\bf x}(\bm\xi,t)$ and the current
density ${\bf j}({\bf x},t)$ are in a one-to-one
correspondence. Therefore our theory can in principle be viewed as a
particular realization of TDCDFT. However it is also legitimate, and
perhaps even more natural, to consider TDDefFT as an independent
member of the family of time-dependent DFT-like theories, such as
TDDFT by Runge and Gross \cite{RunGro1984}, and TDCDFT proposed by
Vignale and Kohn \cite{VigKohn1996}. An apparent advantage of the
deformation-based formalism is the existence of a well founded local
approximation for xc potentials in the KS formulation of the
theory. In fact, TDDefFT provides the most natural and unified
framework for interpreting all currently known local nonadiabatic
approximations
\cite{VigKohn1996,VigUllCon1997,TokPRB2005b,KurBae2004,KurBae2005,TokLNP2006,TaoVig2006}.
The exact representations for the xc potentials derived in Sec.~IV
should result in the further progress in constructing new practical
nonadiabatic functionals.

\appendix
\section{Local symmetries and conservation laws}

In this Appendix we derive two local conservation laws for a $N$-body
system placed in the space with metric $g_{ij}(\bm\xi,t)$, and
subjected to an external field that is generated by the four-potential
$\mathcal A_{0}(\bm\xi,t)$, $\bm{\mathcal A}(\bm\xi,t)$. The dynamics
of the $N$-body wave function
$\Psi({\bm\xi}_{1},\dots,{\bm\xi}_{N},t)$ is governed by the
time-dependent Schr\"odinger equation with the following Hamiltonian
\begin{eqnarray} \nonumber
H &=& \sum_{\alpha = 1}^{N}g^{-\frac{1}{4}}_{\alpha}
[i\partial_{\xi^{i}_{\alpha}} + {\cal A}_{i}(\bm\xi_{\alpha})]
\frac{\sqrt{g_{\alpha}}g^{ij}_{\alpha}}{2m}
[i\partial_{\xi^{j}_{\alpha}} + {\cal A}_{j}(\bm\xi_{\alpha})]
g^{-\frac{1}{4}}_{\alpha} \\
&& \qquad + \sum_{\alpha = 1}^{N}\mathcal A_{0}(\bm\xi_{\alpha})
+ \frac{1}{2}\sum_{\alpha,\beta}
V(l_{\bm\xi_{\alpha}\bm\xi_{\beta}})
\label{A1}
\end{eqnarray}
where $g^{ij}_{\alpha}=g^{ij}(\bm\xi_{\alpha},t)$ and
$l_{\bm\xi_{\alpha}\bm\xi_{\beta}}$ is the length of geodesic that
connects points $\bm\xi_{\alpha}$ and $\bm\xi_{\beta}$. 

Below we derive local balance equations from local 
symmetries of the Dirac-Frenkel action functional (for a similar
derivation see also Ref.~\onlinecite{SonWin2006}):
\begin{equation}
S[\Psi,\mathcal A_{0},\bm{\mathcal A},g_{ij}] = 
\int dt\prod_{\alpha = 1}^{N} d{\bm\xi}_{\alpha}\Psi^{*}\Big(
i\stackrel{\leftrightarrow}{\partial_{t}} - H\Big)\Psi.
\label{A2}
\end{equation}
In particular the local gauge invariance of $S$ is responsible for the
local conservation of the number of particles, while the general
coordinate invariance of the action yields the local momentum balance
equation. 

Let us consider the gauge invariance first. By a direct substitution
we find that the following transformation
\begin{eqnarray}
&& \quad \Psi' = \Psi e^{i\sum_{\alpha}\phi(\bm\xi_{\alpha},t)},
\label{A3}\\ 
\mathcal A'_{0} &=& \mathcal A_{0} - \partial_{t}\phi(\bm\xi ,t),
\quad {\mathcal A}'_{i} = {\mathcal A}_{i} 
+ \partial_{\xi^{i}}\phi(\bm\xi ,t)
\label{A4}\\ 
&& \quad g'_{ij}(\bm\xi ,t) = g_{ij}(\bm\xi ,t),
\label{A5}
\end{eqnarray}
where $\phi(\bm\xi ,t)$ is an arbitrary function, leaves the action
$S$ unchanged, i.~e., 
\begin{equation}
S[\Psi',{\mathcal A}'_{0},\bm{\mathcal A}',g'_{ij}] = 
S[\Psi,\mathcal A_{0},\bm{\mathcal A},g_{ij}].
\label{A6}
\end{equation}
Inserting the infinitesimal version ($\phi\to 0$) of
Eqs.~(\ref{A3})--(\ref{A5}) into Eq.~(\ref{A6}) we get the following condition 
\begin{equation}
\delta S|_{\text{extr}} = \int dt d\bm\xi\left(
-\frac{\delta S}{\delta\mathcal A_{0}}\partial_{t}\phi
+ \frac{\delta S}{\delta{\mathcal A}_{i}}\partial_{\xi^{i}}\phi
\right) = 0
\label{A7}
\end{equation}
As usual we take the variation of the action at the extremal
``trajectory'' that is defined by the equation $\delta
S/\delta\Psi=0$. Therefore the change of the wave function,
Eq.~(\ref{A3}), does not contribute to $\delta S$, Eq.~(\ref{A7}). The
requirement that Eq.~(\ref{A7}) is fulfilled for any $\phi$ yields the
continuity equation 
\begin{equation}
\partial_{t}n + \partial_{\xi^{i}}j^{i} = 0,
\label{A8}
\end{equation}
where the density $n(\bm\xi,t)$ and the current $j^{i}(\bm\xi,t)$ are
defined as follows
\begin{eqnarray}
n = -\frac{\delta S}{\delta\mathcal A_{0}} = \langle\Psi|
\frac{\delta H}{\delta\mathcal A_{0}}|\Psi\rangle,
\label{A9}\\
j^{i} = \frac{\delta S}{\delta\mathcal A_{i}} = -\langle\Psi|
\frac{\delta H}{\delta\mathcal A_{i}}|\Psi\rangle.
\label{A10}
\end{eqnarray}

Similarly the momentum balance equation follows from the invariance of
the action $S$, Eq.~(\ref{A2}), under a general nonsingular
transformation of coordinates, $\bm\xi'=\bm\xi'(\bm\xi,t)$, where
$\bm\xi'(\bm\xi,t)$ is an arbitrary (invertible) function. The
transformation of fields, which leaves the action invariant, is the
following 
\begin{eqnarray}
\Psi'(\{\bm\xi'_{\beta}\}_{\beta =1}^{N},t) 
&=&  \prod_{\alpha = 1}^{N}{\left|
\frac{\partial \bm\xi_{\alpha}}{\partial
\bm\xi'_{\alpha}}\right|}^{\frac{1}{2}}
\Psi(\{\bm\xi_{\beta}\}_{\beta =1}^{N},t), 
\label{A11}\\
 g'_{ij}(\bm\xi',t) &=& 
\frac{\partial\xi^{k}}{\partial\xi'^{i}}
\frac{\partial\xi^{p}}{\partial\xi'^{j}}g_{kp}(\bm\xi,t),
\label{A12}\\
\mathcal A'_{k}(\bm\xi',t) &=& 
\frac{\partial\xi^{i}}{\partial\xi'^{k}}\mathcal A_{i}(\bm\xi,t)
- m g'_{ij}(\bm\xi',t)\frac{\partial\xi'^{i}}{\partial t},
\label{A13}\\
\mathcal A'_{0}(\bm\xi',t) &=& \mathcal A_{0}(\bm\xi,t)
 + \mathcal A'_{i}(\bm\xi',t)
\frac{\partial\xi'^{i}}{\partial t} \nonumber \\
&+& \frac{m}{2}g'_{ij}(\bm\xi',t)
\frac{\partial\xi'^{i}}{\partial t}
\frac{\partial\xi'^{j}}{\partial t},  
\label{A14}
\end{eqnarray}
where $\bm\xi'_{\alpha}=\bm\xi'(\bm\xi_{\alpha},t)$. Let us consider
an infinitesimal transformation of coordinates, which is generated by
the function $\bm\xi' = \bm\xi + \bm\eta(\bm\xi,t)$, $\bm\eta\to
0$. The corresponding change of the action (at the extremal) takes the form
\begin{equation}
\int dt d\bm\xi\left(
\frac{\delta S}{\delta g_{ij}}\delta g_{ij}
+ \frac{\delta S}{\delta\mathcal A_{0}}\delta\mathcal A_{0}
+ \frac{\delta S}{\delta{\mathcal A}_{i}}\delta{\mathcal A}_{i}
\right) = 0
\label{A15}
\end{equation}
In equation (\ref{A15}) $\delta g_{ij}$, $\delta\mathcal A_{0}$, and
$\delta{\mathcal A}_{i}$ are given by the infinitesimal version of
Eqs.~(\ref{A12})--(\ref{A14}):
\begin{eqnarray}
\delta g_{ij} &=& - \eta^{k}\partial_{\xi^{k}}g_{ij}
- g_{ik}\partial_{\xi^{j}}\eta^{k}
- g_{jk}\partial_{\xi^{i}}\eta^{k},
\label{A16}\\
\delta \mathcal A_{i} &=& 
- \eta^{k}\partial_{\xi^{k}}\mathcal A_{i} 
- \mathcal A_{k}\partial_{\xi^{i}}\eta^{k}
- mg_{ik}\partial_{t}\eta^{k},
\label{A17}\\
\delta\mathcal A'_{0} &=& 
- \eta^{k}\partial_{\xi^{k}}\mathcal A_{0}
+ \mathcal A_{k}\partial_{t}\eta^{k}.
\label{A18}
\end{eqnarray}
Inserting Eqs.~(\ref{A16})--(\ref{A18}) into Eq.~(\ref{A15}), and
integrating by parts we obtain at the following condition of the
general coordinate invariance of the action: 
\begin{eqnarray} \nonumber
\partial_{t}\left(mg_{ik}\frac{\delta S}{\delta{\mathcal A}_{i}}
- {\mathcal A}_{k}\frac{\delta S}{\delta{\mathcal A}_{0}}\right)
&+& \partial_{\xi^{i}}\left(2g_{kj}\frac{\delta S}{\delta g_{ij}}
+ {\mathcal A}_{k}\frac{\delta S}{\delta{\mathcal A}_{0}}\right) \\
- \frac{\delta S}{\delta{\mathcal A}_{0}}
\partial_{\xi^{k}}{\mathcal A}_{0} - 
\frac{\delta S}{\delta{\mathcal A}_{i}}
\partial_{\xi^{k}}{\mathcal A}_{i} &-& \frac{\delta S}{\delta g_{ij}}
\partial_{\xi^{k}}g_{ij} = 0.
\label{A19}
\end{eqnarray}
Equation (\ref{A19}) can be further simplified as follows
\begin{eqnarray} \nonumber
m\partial_{t}j_{k} &-& j^{i}(\partial_{\xi^{i}}{\mathcal A}_{k} 
- \partial_{\xi^{k}}{\mathcal A}_{i}) + n\partial_{t}{\mathcal A}_{k}
+ n\partial_{\xi^{k}}{\mathcal A}_{0}\\
&+& \partial_{\xi^{i}}\left(2g_{kj}\frac{\delta S}{\delta g_{ij}}
\right) - \frac{\delta S}{\delta g_{ij}}\partial_{\xi^{k}}g_{ij}=0,
\label{A20}
\end{eqnarray}
where we have used the continuity equation, Eq.~(\ref{A8}), and the
definitions of the density, and the current, Eqs.~(\ref{A9}) and
(\ref{A10}). The last two terms in the left hand side of
Eq.~(\ref{A20}) are easily recognized as a covariant divergence the
following symmetric second rank tensor
\begin{equation}
P^{ij} = \frac{2}{\sqrt{g}}\frac{\delta S}{\delta g_{ij}}
= - \frac{2}{\sqrt{g}}
\langle\Psi|\frac{\delta H}{\delta g_{ij}}|\Psi\rangle.
\label{A21}
\end{equation}
Indeed, using the definition of Eq.~(\ref{A21}) we find
\begin{eqnarray} \nonumber
&&\qquad \partial_{\xi^{i}}\left(2g_{kj}\frac{\delta S}{\delta g_{ij}}
\right) - \frac{\delta S}{\delta g_{ij}}\partial_{\xi^{k}}g_{ij}\\
&=&\sqrt{g} \left(\frac{1}{\sqrt{g}}
\partial_{\xi^{i}}\sqrt{g}P^{i}_{k} 
- \frac{1}{2}P^{ij}\partial_{\xi^{k}}g_{ij}\right)
\equiv \sqrt{g}\nabla_{i}P^{i}_{k}
\nonumber
\end{eqnarray}
Therefore the condition of the general coordinate invariance,
Eq.~(\ref{20}), takes the standard form of the local momentum balance
equation 
\begin{eqnarray} \nonumber
m\partial_{t}j_{k} &-& j^{i}(\partial_{\xi^{i}}{\mathcal A}_{k} 
- \partial_{\xi^{k}}{\mathcal A}_{i}) + n\partial_{t}{\mathcal A}_{k}
\\
&+& n\partial_{\xi^{k}}{\mathcal A}_{0}+ \sqrt{g}\nabla_{i}P^{i}_{k} =0.
\label{A22}
\end{eqnarray}
Apparently the tensor $P^{ij}$ defined after Eq.~(\ref{A21}) plays a
role of the physical stress tensor.



\begin{thebibliography}{37}
\bibitem[{not({\natexlab{a}})}]{note0}
\bibinfo{note}{In the semiclassical limit DFT reduces to the Thomas-Fermi
  theory, which is the hydrostatics of the electron gas.}
\bibitem[{not({\natexlab{c}})}]{note2}
\bibinfo{note}{Throughout this paper we use Latin indexes, $i,j,k,$ etc., to
  label Cartesian axes. Additional Greek indexes will be used in Sec.~III for
  labeling particles in the many-body system. We also employ the standard
  convention of summation over repeated Latin indexes.}
\end{thebibliography}

\end{document}